\documentclass[letterpaper,journal]{IEEEtran}
\usepackage{amsmath,amsfonts}
\usepackage{algorithmic}
\usepackage{algorithm}
\usepackage{array}
\usepackage[caption=false,font=normalsize,labelfont=sf,textfont=sf]{subfig}
\usepackage{textcomp}
\usepackage{stfloats}
\usepackage{verbatim}
\usepackage{graphicx}
\usepackage{cite}
\usepackage{amssymb}
\usepackage{physics}
\usepackage{epstopdf}
\usepackage{cases}
\usepackage{indentfirst}
\usepackage{balance}
\usepackage{booktabs}
\usepackage[hyphens]{url}
\usepackage{breakurl}
\usepackage[thmmarks,amsmath]{ntheorem}
\qedsymbol{\ensuremath{\square}}
\theoremseparator{:}
\newtheorem{theorem}{Theorem}

\usepackage{color}
\theoremsymbol{\ensuremath{\color{Gainsboro}\blacksquare}}
\begin{document}

\title{Control-Oriented Deep Space Communications For Unmanned Space Exploration}

\author{ Xinran~Fang,~\IEEEmembership{Student Member,~IEEE,}
	Wei~Feng,~\IEEEmembership{Senior Member,~IEEE,} Yunfei~Chen,~\IEEEmembership{Senior Member,~IEEE,}
	%	Dingyou~Ma,\\
	Ning~Ge,~\IEEEmembership{Member,~IEEE}, and Gan~Zheng,~\IEEEmembership{Fellow, IEEE}
	%	Yimin~Liu,~\IEEEmembership{Member,~IEEE,}

	\thanks{

	X.~Fang, W.~Feng, and N. Ge  are with the Department of Electronic Engineering, Tsinghua University, Beijing 100084, China (e-mail: {fxr20}@mails.tsinghua.edu.cn, {fengwei}@tsinghua.edu.cn, and  {gening}@tsinghua.edu.cn).

	Y. Chen is with the Department of Engineering, University of Durham, Durham DH1 3LE, U.K. (e-mail: {yunfei.chen}@durham.ac.uk).

	G. Zheng  is with the School of Engineering, University of Warwick, CV4 7AL Coventry, U.K. (e-mail: {gan.zheng}@warwick.ac.uk).
		}

}

%      C.-X. Wang is with the National Mobile Communications Research Laboratory, School of Information Science and Engineering, Southeast University, Nanjing, 210096, China, and also with the Purple Mountain Laboratories, Nanjing 211111, China (e-mail: chxwang@seu.edu.cn).}}

\maketitle
\begin{abstract}
In unmanned space exploration, the cooperation among  space robots requires advanced communication techniques. In this paper, we propose a communication optimization scheme for a specific cooperation system named the ``mother-daughter system”. In this setup, the mother spacecraft orbits the planet, while daughter probes are distributed across the planetary surface. During each control cycle, the mother spacecraft senses the environment, computes control commands and distributes them to daughter probes for actions. They synergistically form sensing-communication-computing-control ($\mathbf{SC^3}$) loops. Given the indivisibility of the $\mathbf{SC^3}$ loop, we optimize the mother-daughter downlink for closed-loop control.  The optimization objective is the  linear quadratic regulator (LQR) cost, and the optimization parameters are the block length and transmit power. To solve the nonlinear mixed-integer problem, we first identify the optimal block length and then transform the power allocation problem into a tractable convex problem. We further derive the approximate closed-form solutions for the proposed scheme and two communication-oriented schemes: the max-sum rate scheme and the max-min rate scheme. On this basis, we analyze their power allocation principles. In particular, for time-insensitive control tasks, we find that the proposed scheme demonstrates equivalence to the max-min rate scheme. These findings are verified through simulations.
\end{abstract}


\section{Introduction}
\subsection{Background and Motivation}
Communication plays a vital role in deep space exploration. The link between the spacecraft and the earth command center is the lifeline of the spacecraft.  This connection provides tracking, telemetry and command (TT\&C) services, which ensure the correct operation.  According to the definition of the International Telecommunication Union , if the spacecraft-Earth distance is greater than $2\times10^6$ km, the communication between the spacecraft and the earth belongs to deep space communication (DSC) \cite{ref111}. Due to the great distance, DSC exhibits unique characteristics. The first is the large latency. Taking Mars as an example, the propagation delay from the perigee of the Mars to the Earth is 4 minutes, while from the apogee of the Mars, it is 24 minutes \cite{ref112}. The second is the low data rate. As a common sense, the signal power decays at least quadratically with distance.
To ensure the same strength of the received signals, the transmit power of the spacecraft from Mars should be $9\times10^{16}$ times stronger than that of a smart phone located 1 km away from the receiver \cite{ref112}. However, constrained by size and weight limitations, the on-board base station (BS) carried by the spacecraft is less powerful than ground-based facilities. It can only transmit signals at very low power. Ground receivers must be equipped with large aperture antennas to capture the weak signals from space. Even so, the data rate between the spacecraft and the earth command center remains quite low, e.g., several megabits per second in NASA's Deep Space Network (DSN) \cite{ref114}.

With the growing curiosity on outer space, current space activities touch the fundamental problems of DSC. The sheer number of space missions is one factor. According to the United Nations, the number of space robots has grown exponentially in the last decade \cite{ref115}. With advanced sensing techniques, new spacecrafts need a higher data rate to send high-resolution data back, which dramatically increases the burden of the DSN. Moreover, in addition to observation-based explorations, major space powers like the United States, the European Union and China have announced plans to launch execution-based missions such as sample return and base constructions. Unmanned space exploration is evolving from the one that is characterized by an individual spacecraft to the one that relies on network operations. This, however, exceeds the competence of current DSNs.

Unlike space-ground communication that is suffered from long distances, the space-space communication enables fast and flexible connections among space robots, fostering robotic cooperation. With the connection of the communication links, space robots operate cyclically: sensors collect surrounding information and report it to the computing unit, the computing unit analyzes the situation and distributes calculated commands to actuators, and the actuators take actions. These robots synergistically form sensing-communication-computing-control ($\mathbf{SC}^3$) loops.
However, communication often becomes the critical bottleneck in the $\mathbf{SC}^3$ loop.  Firstly, to deal with the uncertain environment, the $\mathbf{SC}^3$
loops run in small cycle time. This constrains the time for data transmissions. When the spacecrafts use the short packet to transmit information, the achievable  rate suffers significant degradation compared with the traditional Shannon capacity.  In addition, the BSs carried by the spacecraft are less powerful compared with ground facilities. This further limits the communication data rate. To ensure that communication enhances rather than constrains unmanned space exploration, addressing communication challenges emerges as a pivotal factor in enhancing the $\mathbf{SC}^3$ loop.
%The formed cooperation system exhibits prominent task-oriented characteristics. In most cases, space robots operate cyclically: sensors collect surrounding information and report it to the computing unit, the computing unit analyzes the situation and distributes calculated commands to actuators, and the actuators take actions.
%This is quite different from traditional communication networks.

\begin{figure*}
	\centering
	\includegraphics[width=0.85\textwidth]{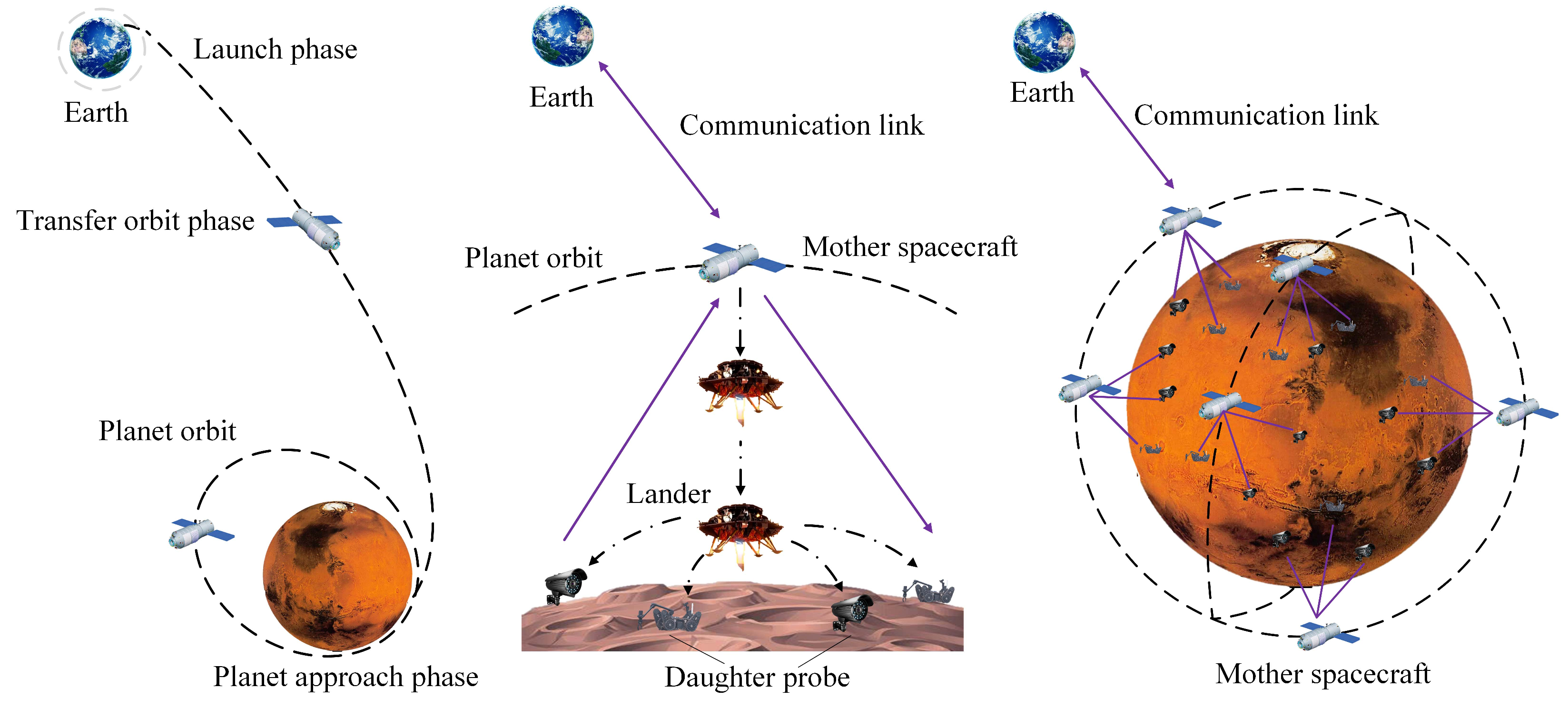}
	\caption{Illustration of the ``mother-daughter system" in unmanned space exploration.}
	\label{fig.1}
\end{figure*}

In this paper, we explore the communication scheme for a cooperation system named the ``mother-daughter system". As the name implies, this system comprises a central robot serving as the ``mother", alongside multiple less advanced robots functioning as the ``daughters". To facilitate daughter robots’ tasks, the mother robot diligently oversees daughter robots and provides them with sensing and computing assistance and  operational guidelines. In fact, the ``mother-daughter system" is not a new concept. We are able to find many prototypes of such systems in current space missions. In Mars 2020, the Perseverance rover (``mother") and the Ingenuity helicopter (``daughter") is a typical example. Throughout Ingenuity's missions, Perseverance monitored its flight and relayed data between the Ingenuity and the Earth control center. Supported by the Perseverance, the Ingenuity navigated numerous unforeseen challenges beyond its initial design, including choosing landing sites in treacherous terrain, dealing with a dead sensor, and cleaning itself after dust storms \cite{bb1}. Another example is the Lunar gateway, which is a vital component of the NASA-led Artemis missions planned to be launched in 2025.  This space station will be a multi-purpose outpost orbiting the Moon, acting as a ``mother" space station to support ``daughter" landers, rovers, and human missions on the Moon's surface.
Such configurations, involving a central spacecraft (``mother"), coordinating with multiple surface or subsurface probes (``daughters"), are increasingly recognized as a basic coordination pattern for the unmanned space exploration.
In Fig. \ref{fig.1}, we illustrate the envisaged ``mother-daughter system" in  future unmanned space missions. Upon the launch of a space mission, the mother spacecraft progresses through launch, transfer orbit, and planet approach phases to reach its designated orbit. It then releases the lander, which descends to the planetary surface and releases daughter probes. To enable these less advanced probes to conduct exploration tasks in varying environments, the mother spacecraft provides real-time guidelines to them. Together, they form $\mathbf{SC}^3$ loops, serving as the fundamental unit for task execution. In addition, the adaptable orchestration of the ``mother-daughter system" can further give rise to a large-scale exploration network.

In the literature, few studies have investigated space communication from the perspective of the $\mathbf{SC}^3$ loop. Communication researchers mostly took the communication link as the investigated objective and addressed the unique challenges in DSC \cite{f50}. While improving communication performance is crucial for space exploration, separating the communication process from the $\mathbf{SC^3}$ loop might not be optimal. This approach overlooks the communication impact on the overall functioning of the $\mathbf{SC^3}$ loop. In contrast, researchers in control have studied the whole $\mathbf{SC^3}$ loop. Specifically, in the wireless control system (WCS) \cite{f29,f30,f31}, which is a type of distributed internet of things (IoT) system where sensors, controllers, and actuators are interconnected via a wireless network, communication researchers began to take the $\mathbf{SC^3}$ loop as the research objective and proposed the control-aware scheme. In the following, we review related studies of DSC and WCS.

\subsection{Related Works}
\subsubsection{DSC}
DSC plays an important role in unmannd space exploration. Due to the long transmission distance, there exist unique challenges for DSC. Related studies have addressed different challenges from different perspectives, including communication protocols, networking, and resource optimization. For example, Ha \emph{et al.} proposed a reinforcement learning-based link selection strategy to improve the streaming performance of DSC \cite{ff2}. To tackle the solar scintillation effect, Xu \emph{et al.} devised a dual-hop mixed communication system, which comprises a radio frequency link between the Earth and the relay satellite and a free-space optical  link between the satellite and a deep space probe \cite{ff3}. To address the challenge of intermittent connections, the delay-tolerant network (DTN) has been developed, in which the store-and-forward strategy is used to combat the interruption. On this basis, Yang \emph{et al.} proposed a hybrid bundle retransmission scheme to ensure the DTN reliability \cite{ff4}. Rango \emph{et al.} proposed an adaptive bundle rate scheme to tackle concurrent bundle transmissions \cite{ff5}. In terms of the network structure, Wan \emph{et al.} proposed a satellite relay constellation network and detailed the mathematical model, topology design and performance analysis \cite{ff6}.

In addition to these straightforward communication schemes, some cutting-edge technologies can further enhance the deep space network.
For instance, the non-orthogonal multiple access (NOMA) technique can be introduced to allow multiple space robots to share the same spectrum resources. Its application in DSC could enable more efficient usage of spectrum resources.
Moreover, equipping space robots with advanced and compact hardware, such as reconfigurable pixel antennas and integrated communication and sensing platforms, could enhance their operational capabilities.
Furthermore, the age of information (AoI) provides a new perspective to measure the freshness of data.
Recently, Yuan \emph{ et al.} proposed an AoI-guided user scheduling scheme for the communication between the Lunar far surface and the Earth. The proposed scheme was proven effective to ensure the timely delivery of sensing data \cite{bb2}. Thinking more broadly, the unmanned aerial vehicle (UAV) can be uploaded to the explored planet and establish the satellite-UAV-ground network to felicitate large-scale unmanned space exploration \cite{ff7}\cite{ff8}. These techniques provide valuable insights for enhancing the deep space network and deserves more research attentions.

\subsubsection{WCS}
Taking communication imperfections, such as delays, dropouts, and rate limitations into account, researchers in control have paid great attention to the relationship between communication and control.
%, e.g. stability analysis and control performance limit analysis.
For example, Tatikonda \emph{et al.} investigated the minimal data-rate requirement to stabilize the linear time-invariant system \cite{f33}. It was proven that only when the data rate exceeds the intrinsic entropy, can the system  be stabilized \cite{f34}. Afterward, Nair \emph{et al.} extended this result to nonlinear systems \cite{f35}. Kostina \emph{et al.} further generalized these results to vectorial, non-Gaussian, and partially observed systems. Given the data rate, the authors derived the lower bound of the control linear quadratic regulator (LQR) cost \cite{f12}. %Branicky \emph{et al.} took the constant packet delay into account and found that the stable condition is determined by the Schurness of the quantization-delay matrix \cite{f11}.
%Afterward, the stable condition was investigated under different communication delay \cite{f11, f14,f15}.

Building upon these results, numerous control-aware schemes have been proposed for WCSs. Weiner \emph{et al.} proposed a control-aware wireless system and designed the transmission schemes in physical and media access control layers  to guarantee  the latency of the command transmission \cite{j0}.
Gatsis \emph{et al.} proposed a sensor transmission policy to minimize the LQR cost while constraining of the long-term communication resource consumption \cite{j1}.
Lyu \emph{et al.}
devised a control-aware cooperative transmission scheme that jointly optimized  channel allocation, power control, and relay cooperation \cite{ref9}. Tong \emph{et al.} considered a predictive-based control system, which transmits a set of predicted commands in a single transmission. The authors optimized the predictive  command length to minimize the communication consumption \cite{j3}. Girgis \emph{et al.} provided a prediction-based control scheme to minimize the AoI and transmit power \cite{ref10}.
Chang \emph{et al.} optimized the bandwidth, transmit power, and control convergence rate to maximize the spectrum efficiency \cite{ref13}. Yang \emph{et al.} developed a framework to optimize a new metric named the energy-to-control efficiency \cite{ref11}. Lei \emph{et al.} investigated a satellite-UAV system and proposed a multi-loop optimization scheme to minimize the sum LQR cost. The transmit power was balanced among different loops \cite{ref21}. Esien \emph{et al.} proposed a control-aware scheduling scheme that adapted transmissions to channel dynamics and system states \cite{ref20}. Recently, Ali \emph{et al.} incorporated  the deep learning approach to optimize the sampling period, communication block length, and packet error probability to minimize the energy consumption \cite{j2}.
While these works took the first step to investigate the control-aware communication design, challenges remain in the considered space setting.
Firstly, the space environment is totally uncertain. The $\mathbf{SC}^3$ loop needs to run in small cycle time to adapt to the varying environments. This restricts the time window of data transmissions.
When using short packets in communication, the achievable  rate in the finite block length (FBL) regime is largely reduced compared with the Shannon capacity.  Moreover, given the payload constraint, space robots are equipped with lightweight transceivers, whose communication capabilities is far away from ground-based facilities. The time sensitivity of the $\mathbf{SC}^3$ loop and limited communication resources of the space robots call for the accurate modeling and efficient usage of communication in space missions.
Current studies either failed to prioritize control metrics as their objectives or did not model the data transmissions in the FBL regime, making them lose efficiency in the space environment. Tab. \ref{tab2} presents a qualitative comparison of this work and related work in WCSs.

\begin{table*}
	\caption{Quantitative Comparisons between the proposed scheme and the related works in WCSs}
	\label{tab2}
 \begin{tabular}{|p{2.3cm}<{\centering}|p{4cm}<{\centering}|p{4.2cm}<{\centering}|p{2.5cm}<{\centering}|p{2cm}<{\centering}|}
	\toprule
	\bf{Work}&\bf{System Setting}&\bf{Optimization Objective}&\bf{Short-packet Transmission}&\bf{Loop Number} \\
	\cline{1-5}
	Weiner \emph{et al.} \cite{j0}&Real-time WCS&Communication: latency&Yes&Multiple loops\\
	\cline{1-5}
	Gatsis \emph{et al.} \cite{j1}&General WCS &Control: LQR cost &No &Single loop\\
	\cline{1-5}
	Lyu \emph{et al.} \cite{ref9}&Multi-hoop WCS &Hybrid: transmission ability and the upper bounds of the LQR cost&No &Multiple loops\\
	\cline{1-5}
	Tong \emph{et al.} \cite{j3}& Pridiction-based WCS
	&Communication: energy consumption &No & Single loop\\
	\cline{1-5}
	Girgis \emph{et al.} \cite{ref10}&General WCS&Hybrid: a weighted summation of the AoI and the transmission power&No&Multiple loops\\
	\cline{1-5}
	Chang \emph{et al.} \cite{ref13}&Real-time WCS&Communication: energy consumption&Yes&Single loop\\
	\cline{1-5}
	Yang \emph{et al.} \cite{ref11}&Real-time WCS&Hybrid: energy-to-control efficiency&Yes&Multiple loops\\
	\cline{1-5}
	Lei \emph{et al.} \cite{ref21}&Satellite-UAV system&Control: limit LQR cost&No&Multiple loops\\
	\cline{1-5}
	Esien \emph{et al.} \cite{ref20}&Real-time WCS& Communication: transmission time &Not covered &Multiple loops\\
	\cline{1-5}
	Ali \emph{et al.} \cite{j2}&Real-time WCS &Communication: power consumption&Yes& Multiple loops\\
	\cline{1-5}
	This work&Mother-daughter system in space&Contol: limit LQR cost&Yes&Multiple loops\\
	\bottomrule
\end{tabular}
\end{table*}

Recently, there are some works that considered the joint communication and sensing/computing model in the closed-loop control.  Han \emph{et al.} introduced the reconfigurable intelligence surface to cover the blind areas of the WCS. The authors introduced the rate-distortion function to model the sensing phase, which imposes the uplink to transmit a certain quantity of data to meet the distortion requirement. The proposed scheme remained a communication scheme that jointly optimized the transmission power, transmission time, beamforming and the reflecting coefficients of RIS \cite{b2}. Li \emph{et al.} proposed a joint communication and computing optimization scheme, which took the communication power and computing frequency as the optimization variable and minimized the closed-loop latency \cite{b1}. These works made initial attempts to combine sensing, communication, and computing within the $\mathbf{SC^3}$ loop modeling. However, a foundational model that can uniformly characterize the relationships among sensing, communication, computation, and control is still unknown. In the considered space setting, we identify communication as the critical bottleneck affecting the overall performance of the $\mathbf{SC^3}$ loop. Given the limited communication capabilities of space robots, maximizing the communication efficiency for the task becomes crucial in enhancing the $\mathbf{SC^3}$ loop. Therefore, our paper focuses on  the communication resource optimization for the closed-loop control. We take the joint sensing, communication, and computing design as an important direction for future research.
%while can not  is not the optimal solution for the assumed task.
% while others ignored the periodic nature of the closed-loop communication. They modeled the communication process under the traditional Shannon framework, overlooking the rate drop resulting from short-packet transmissions.

%The cycle rate is modeled in the FBL regime.

\subsection{Main Contributions}
In this paper, we investigate the communication scheme for the $\mathbf{SC^3}$ loop within the ``mother-daughter system". Given the task-oriented trait of the $\mathbf{SC^3}$ loop, we propose a control-oriented communication scheme, whose distinctive feature is the integration of control objective into the communication design. The main contributions are listed as follows.

\begin{enumerate}
	\item We propose a control-oriented communication scheme for the $\mathbf{SC^3}$ loop within the ``mother-daughter system". Considering the cycle-time constraint, we model the mother-daughter downlink in the FBL regime. On this basis, we optimize the downlink block lengths and transmit power to minimize the sum LQR cost of the ``mother-daughter system".

	\item To solve the nonlinear integer problem, we first derive the optimal block lengths based on the monotonicity of the rate-cost objective function. Subsequently, we rigorously prove the monotonicity and concavity-convexity of the achievable rate expression and the rate-cost function. Building upon these insights, we transform the power allocation problem into a tackle convex problem.

	\item Leveraging the tight approximations of the cycle rate and rate-cost functions, we derive the approximate closed-form solution of the transmit power. We compare this closed-form solution with that of the max-sum rate scheme and the max-min rate scheme. Through comparative analysis, we differentiate the proposed scheme with communication-oriented schemes and find its fairness-mined nature.

	\item We use appropriate parameters to stimulate the space environment and test the proposed scheme. Simulation results confirm the effectiveness of the proposed scheme in improving the control performance of the $\mathbf{SC^3}$ loops and  validate our findings.
\end{enumerate}

\subsection{Organization and Notation}
The rest of this paper is organized as follows. Section \ref{section 2} introduces the model of the ``mother-daughter system" and the $\mathbf{SC^3}$ loop. Section \ref{section 3} presents the  control-oriented communication optimization scheme and its solution. Section \ref{section 4} derives the closed-form solution of the proposed scheme and compares it with that of the max-sum rate scheme and max-min rate scheme.  %We also derive the closed-form solutions of these three schemes, and show the  control-oriented scheme can be approximated into a modified max-min rate scheme in this section.
Section \ref{section 5} presents simulation results.  Section \ref{section 6} draws conclusions.

Throughout this paper, vectors and matrices are represented by lower and upper boldface symbols. $\mathbb{R}^{m \times m}$ represents the set of $m\times m$ real matrices, $\mathbf{I}_m$ is the $m\times m$ unit matrix, and $\mathbf{0}_m$ is the $m\times m$ zero matrix. $\lambda(\mathbf{A})$ denotes the eigenvalue of matrix $\mathbf{A}$.

\section{System Model}
\label{section 2}
\begin{figure}[t]
	\centering
	\includegraphics[width=3.2 in]{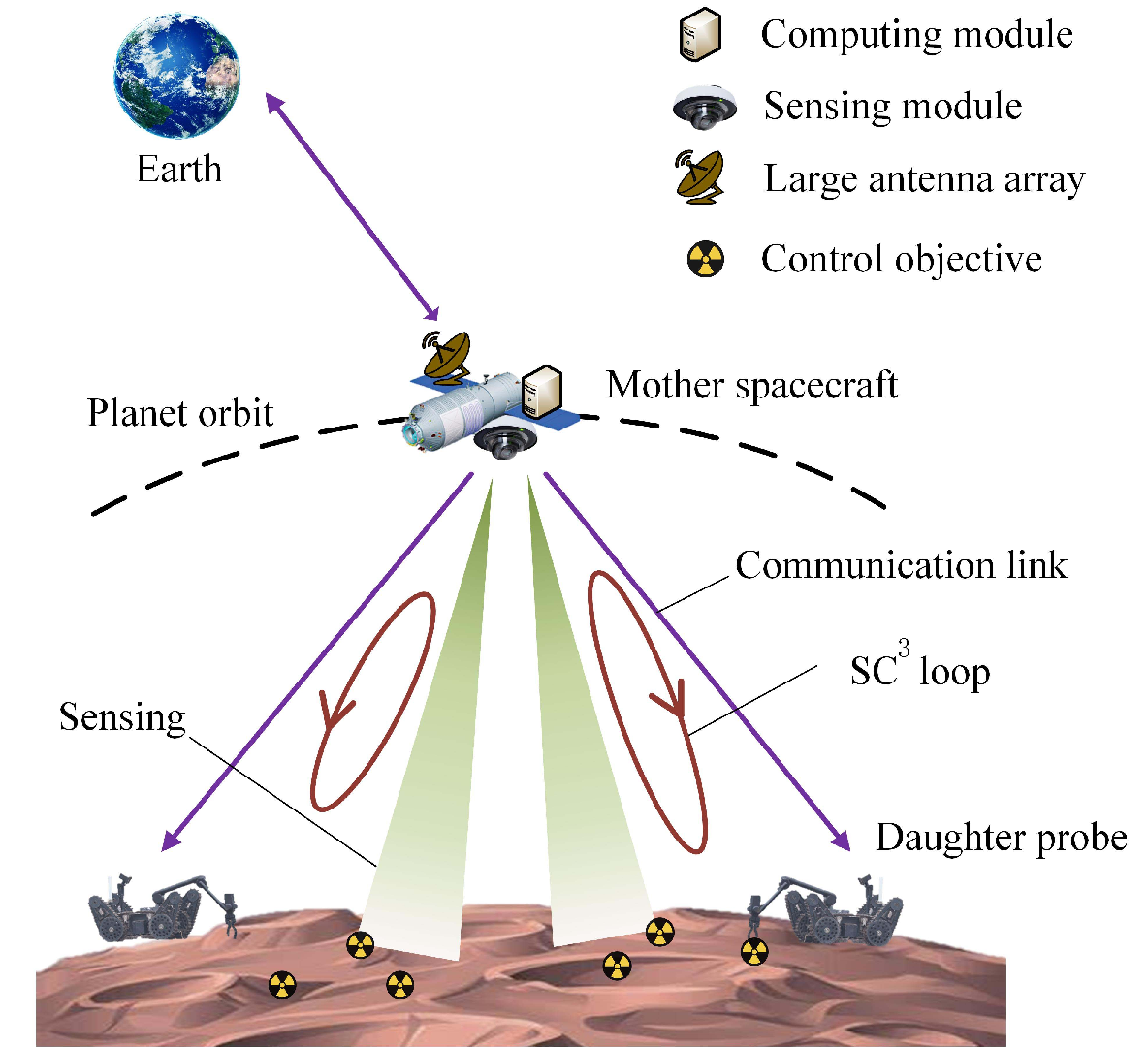}
	\caption{Illustration of an unmanned space cooperation system that comprises a mother spacecraft and multiple daughter probes. The spacecraft and probes synergistically form $\mathbf{SC^3}$ loops, and each  $\mathbf{SC^3}$ loop takes charge of a  control-type subtask. }
	\label{model}
\end{figure}
In Fig. \ref{model}, we present the model of the ``mother-daughter system". The mother spacecraft is located in the planet's orbit and carries advanced sensing, communication, and computing modules.
In each control cycle, it senses system states, calculates control commands and distributes these commands to daughter probes for actions. Working in a cyclical manner, they synergically form multiple $\mathbf{SC^3}$ loops. The mother spacecraft also maintains a communication link with the Earth, which serves for TT\&C purposes and sends crucial data back to the Earth.
%While the "mother-daughter system" setup shares intuitive similarities with Internet of Things (IoT) scenarios, our "mother-daughter system" faces unique challenges operating in the space environment, i.e., the varying environment and restricted communication resources. The varying environment requires the SC loop to run in small cycle time. This imposes strict latency requirement for the data transmissions. The restricted communication resource limits the data rate of the  mother-daughter links. The
We assume that this ``mother-daughter system" includes $K$ $\mathbf{SC^3}$ loops.  Each $\mathbf{SC^3}$ loop operates an LQR-based controller to execute a control-type task.
For the $k$ th $\mathbf{SC^3}$ loop at time index $i$, we use the following linear time-invariant equation to model the dynamics of the controlled system,
\begin{equation}
\label{sta_elv}
\mathbf{x}_{k,i+1} = \mathbf{A}_{k}\mathbf{x}_{k,i}+\mathbf{B}_{k}\mathbf{u}_{k,i}+\mathbf{v}_{k,i},
\end{equation}
where $\mathbf{x}_{k,i}\in\mathbb{R}^{m\times1}$  represents the system state, $\mathbf{u}_{k,i}\in\mathbb{R}^{m\times1}$ represents the control actions, and
$\mathbf{v}_{k,i}\in\mathbb{R}^{m\times1}$ represents the system uncertainty, whose covariance matrix is denoted as $\mathbf{\Sigma}_{\mathbf{v}_k}\in\mathbb{R}^{m\times m}$.
The matrix $\mathbf{A}_{k}\in\mathbb{R}^{m\times m}$ is the system dynamics matrix, which quantifies the intrinsic behavior of the states in the absence of control inputs.
The matrix $\mathbf{B}_{k}\in\mathbb{R}^{m\times m}$ signifies the control input, delineating how control actions ($\mathbf{u}_{k,i}$) influence the system state transition from $\mathbf{x}_{k,i}$ to $\mathbf{x}_{k,i+1}$. Taking the temperature control as an example, the system state is thus the temperature.
$\mathbf{A}_{k}$ models how the current temperature evolves into the next value without outside intervention and $\mathbf{B}_{k}$ models the impact of the control action on the temperature changes.
Back to our setting, we assume the system is unstable $(\lambda(\mathbf{A}_k)>1)$, and  it can be stabilized using $(\mathbf{A}_k,\mathbf{B}_k)$.
In practice, the state-space equation is non-linear. \eqref{sta_elv} is obtained by linearizing the system state around the working point. Currently, the LQR controller has been widely applied in space exploration, for tasks like spacecraft maneuvering and formation flying \cite{ref53}. Although simplified, the linear control model is effective in characterizing the system dynamics.

During each control cycle, the spacecraft first senses system states. The sensing process is modeled as a partially observed process:
\begin{equation}
\mathbf{y}_{k,i}=\mathbf{C}_{k}\mathbf{x}_{k,i}+\mathbf{w}_{k,i},
\end{equation}
where $\mathbf{y}_{k,i}\in\mathbb{R}^{m\times1}$ is the observation state, $\mathbf{C}_{k}\in\mathbb{R}^{m\times m}$ is the observation matrix, and $\mathbf{w}_{k,i}\in\mathbb{R}^{m\times1}$ is the sensing noise, which conforms to the Gaussian distribution.  The covariance matrix  of $\mathbf{w}_{k,i}$ is denoted as $\mathbf{\Sigma}_{\mathbf{w}_k}=\sigma^2_w\mathbf{I}_m$.

Then, the spacecraft processes sensing data and calculates control commands. Since the LQR-based control is grounded in  well-established control theory, the optimal control input, i.e., $\mathbf{u}_{k,i}$, can be calculated by solving a well-defined optimization problem \cite{ref113}. %In this sense, the computing process does not introduce performance degradation on the closed-loop control.

Afterward, the mother spacecraft functions as the BS, sending control commands to the respective probes. We assume that both the mother spacecraft and daughter probes are equipped with a single antenna.
This single-antenna setting can be  extended to the multi-antenna setting. Notably, since the channel in space environment is less affected by small-scale fading, the multi-antenna configuration is utilized to implement beamforming. By concentrating the signal energy towards the intended receiver, beamforming significantly increases the signal-to-noise ratio (SNR). This improvement in the SNR is crucial for our ``mother-daughter system", where the signal strength can be severely attenuated due to the vast distance of the mother-daughter downlink. In addition to enhancing the SNR, directional communication helps minimize inter-link and inter-system interference. This ensures the performance of the whole space network, accommodating the demands of large-scale exploration in the future. We take the control-oriented beamforming as an interesting topic for further research and we do not delve into this topic in this paper. The spacecraft uses orthogonal frequency bands to communicate with probes, allowing different $\mathbf{SC}^3$ loops to operate simultaneously without co-channel interference. The bandwidth is the same for $K$ loops, which is denoted as $B$.  As for the channel model, the near-planet link has similar characteristics as the near-Earth link \cite{f19}. We thereby adopt the model of the near-Earth link \cite{ff81} to characterize the near-planet environment:
\begin{equation}
\label{channel1}
\begin{aligned}
PL[dB]&=P_{Los}[FSPL+SF_1]\\
&\quad+(1-P_{Los})[FSPL+SF_2+CL],
\end{aligned}
\end{equation}
where $PL$ is the path loss, $P_{Los}$ is the line-of-sight probability, $SF_1$ and $SF_2$ are the fading margins reserved for the shadowing fading, and $CL$ is the clutter effect. The specific values of these  parameters depend on the planetary conditions, such as the presence of atmosphere and surface dust. %We neglect the atmospheric effect by assuming that the atmosphere of the explored planet, such as Mars, is much thinner than that of Earth \cite{f19}.
$FSPL$ represents the free-space path loss, which is a function of the carrier frequency $f_c$ (MHz) and the mother-daughter distance $d_k$ (km):
\begin{equation}
FSPL(d_k,f_c)\triangleq32.45+20\log_{10}(f_c)+20\log_{10}{(d_k)}.
\end{equation}
In addition, we calculate the antenna gain of the mother spacecraft \cite{ff11}:
\begin{equation}
\label{eq5}
G_k=G_{\max}\big(\frac{J_1(u_k)}{2u_k}+36\frac{J_3(u_k)}{u_k^3}\big)^2,
\end{equation}
where  $J_1(\cdot)$ and $J_3(\cdot)$ are the first-kind Bessel functions of orders $1$ and $3$, $G_{\max}$ is the maximum antenna gain, and $u_k$ is given by
\begin{equation}
u_k=\frac{2.07123\sin\theta_k}{\sin\theta_{\text{3dB}}},
\end{equation}
where $\theta_{\text{3dB}}$ is the one-sided half-power beam width and $\theta_k$ is the off-axis angle. %Considering the communication happens when the spacecraft is at the top of the daughter probes, the channel condition is
Based on \eqref{channel1} and \eqref{eq5}, the channel gain $g_k$ of the mother-daughter downlink is given by
\begin{equation}
g_k=G_k10^{\frac{-PL[dB]}{10}}.
\end{equation}

In the considered scenario, the spacecraft must send commands to the probes within the cycle time.  These commands need to be encoded by short channel codes, which renders rate losses. According to \cite{ref0}, the achievable rate in the FBL regime can be approximated into
\begin{small}
\begin{equation}
\label{ro}
r(n_k,p_k)\approx \log_2(1+\frac{g_kp_k}{\sigma^2})-\sqrt{\frac{V(p_k)}{n_k}}Q^{-1}(\epsilon_k) \quad (\text{bit/channel use}),
\end{equation}
\end{small}
where $n_k$ is the block length,  $p_k$ is the transmit power, $\log_2(1+\frac{g_kp_k}{\sigma^2})$ is the Shannon capacity, and $\epsilon_k$ is the transmit error probability.   $Q^{-1}(\cdot)$ is the inverse of the $Q$ function, i.e., $Q(x)=\frac{1}{\sqrt{2\pi}}\int_{x}^{\infty}e^{-\frac{t^2}{2}}dt$, and $V(p_k)$ is the channel dispersion, which is approximated by the following expression \cite{ref1}
\begin{equation}
V(p_k)\approx(\log_2e)^2\big(1-\frac{1}{(1+\frac{g_kp_k}{\sigma^2})^2}\big).
\end{equation}
%	When the block length $n_k$ is of tens to hundreds, the uncertain term $O(\frac{\log n_k}{n_k})$  is negligible.
In information theory, one channel use corresponds to transmitting one complex symbol. Therefore, the block length $n_k$  serves as the number of transmit symbols and the channel use. Based on \eqref{ro}, when using a $n_k$-length packet to send the command, the transmit bits are calculated by the following expression:
\begin{small}
\begin{equation}\label{r}
\begin{aligned}
R(n_k,p_k)\approx(1-\epsilon_k)\big(n_k\log_2(1+\frac{g_kp_k}{\sigma^2})-\sqrt{n_kV(p_k)}Q^{-1}(\epsilon_k)\big),
\end{aligned}
\end{equation}
\end{small}
where we denote $R(n_k,p_k)$ as the cycle rate, which represents the transmit information per cycle.  Obviously, a higher cycle rate facilitates the transmission of more precise control commands, leading to better control performance.

Upon receiving the control command, the probe takes control actions, leading the system to evolve as described by \eqref{sta_elv}.
To ensure that the $\mathbf{SC^3}$ processes are finished within the cycle time, denoted as $T_k^0$,  it is necessary to satisfy:
\begin{equation}
\label{k1}
t_k^s+t_k^{c}+t_k^{d}+t_k^{a}\leqslant T_{k}^{0},
\end{equation}
where $t_k^s$, $t_k^{c}$, $t_k^{d}$,  and $t_k^{a}$ are the time for sensing, computing, downlink communication, and executing control actions.  Since we focus on the communication process, the time for sensing, computing, and executing control actions is predetermined in this paper. The communication time includes two parts,
\begin{equation}
t_k^{d}=\frac{n_k}{B}+\tau_k, n_k\in \mathbb{N}^+,
\end{equation}
where $\frac{n_k}{B}$ is the transmission delay and $\tau_k$ is the  propagation delay, which is determined by the mother-daughter distance $d_k \ (\text{m})$,
\begin{equation}
\tau_k=\frac{d_k}{c},
\end{equation}
where $c \ (\text{m/s})$ is the speed of light. We simplify the cycle-time constraint \eqref{k1} into a constraint of the block length
\begin{equation}
\label{ct}
\frac{n_k}{B} \leqslant T_k, \ n_k\in\mathbb{N}^+,
\end{equation}
where
\begin{equation}
\label{rv1}
T_k\triangleq T_{k}^0-t_k^s-t_k^{c}-t_k^{a}-\tau_k,
\end{equation}
which represents the maximum allowable time for the transmission.

% where	$T_k\triangleq T_{k}^0-t_k^S-t_k^{CP}-t_k^{A}$.

Moving forward, we use the LQR cost to measure the control performance of the $\mathbf{SC}^3$ loops. It is a weighted quadratic summation of the state derivations and the control inputs,
\begin{equation}
l_k = \limsup\limits_{N\rightarrow \infty}\mathbb{E} \left[ \sum_{i = 1}^{N} \left(\mathbf{x}_{k,i}^\text{T}\mathbf{Q}_{k}\mathbf{x}_{k,i} +\mathbf{u}_{k,i}^\text{T}\mathbf{R}_{k}\mathbf{u}_{k,i}\right) \right],
\end{equation}
where $\mathbf{Q}_{k}$ and $\mathbf{R}_{k}$ are weight matrices. $\mathbf{Q}_{k}$ quantifies the importance of maintaining state variables within desired values and $\mathbf{R}_{k}$ quantifies the relative cost or undesirability of utilizing control inputs to influence the system state. As a result, the larger $\mathbf{Q}_{k}$ leads to more powerful control actions, while the larger $\mathbf{R}_{k}$ leads to more cost-friendly control actions. %In particular, when $\mathbf{R}_{k}=0$, the control focus is state deviations while not concern about the cost of the control inputs. Under this setting, the LQR cost quantifies the state deviations and reflects the control effectiveness.
In practical implementations, engineers adjust $\mathbf{Q}_{k}$ and $\mathbf{R}_{k}$ to produce a controller consistent with designed goals.
In the context of our ``mother-daughter system", the LQR cost represents a theoretical measurement of the control efficiency, with a lower LQR cost denoting more effective and efficient control of the $\mathbf{SC}^3$ loop.  When it is applied in practical applications, the LQR cost is mapped to a specific physical meaning.  For instance, in regulating extreme temperatures, which is the basis for many  space experiments, e.g., space breeding, the LQR cost quantifies temperature deviations from the desired range and the cost of control inputs. A lower LQR cost represents the temperature is more effectively maintained in the desired interval. Thus, compared with general communication metrics, e.g., rate, the LQR cost has more specific physical meanings and can directly reflect task performance. This is why we take  the LQR cost as the objective in this paper.

According to \cite{f34}, the cycle rate needs to satisfy the following stable condition such that the controlled system can be stabilized,
\begin{equation}\label{min_R}
R(n_k,p_k)  \geqslant \log_2 |\det \mathbf{A}_k|, \ \forall k,
\end{equation}
where $\log_2 |\det \mathbf{A}_k|$ is the intrinsic entropy, which quantifies the system instability level. On the premise that \eqref{min_R} is satisfied, the finite cycle rate constructs a lower bound of the LQR cost \cite{f12}:
\begin{equation}\label{R1}
\begin{aligned}
l_k \geqslant &\frac{m |\det \mathbf{N}_k\mathbf{M}_{k}|^\frac{1}{m}} {2^{\frac{2}{m}(R(n_k,p_k)-\log_2|\det \mathbf{A}_k|)}-1}\\
&\quad+\text{tr}\left( \mathbf{\Sigma}_{\mathbf{v}_k}\mathbf{S}_k\right)+\text{tr}(\mathbf{\Sigma}_k\mathbf{A}_k^T\mathbf{M}_k\mathbf{A}_k),
\end{aligned}
\end{equation}
where
\begin{subequations}\label{Riccati}
\begin{align}
\mathbf{S}_k & = \mathbf{Q}_k + \mathbf{A}_k^\text{T}\left(\mathbf{S}_k- \mathbf{M}_k\right) \mathbf{A}_k,\\
\mathbf{M}_k & = \mathbf{S}_k^T \mathbf{B}_k \left( \mathbf{R}_k + \mathbf{B}_k^T \mathbf{S}_k \mathbf{B}_k\right) ^{-1} \mathbf{B}_k^\text{T} \mathbf{S}_k,\\
\mathbf{N}_k & = \mathbf{A}_k\mathbf{\Sigma}_k\mathbf{A}_k^T-\mathbf{\Sigma}_k+\mathbf{\Sigma}_{\mathbf{v}_k}.
\end{align}
\end{subequations}
$\mathbf{\Sigma}_k$ is calculated by
\begin{equation}
\mathbf{\Sigma}_k=\mathbf{P}_k-\mathbf{P}_k\mathbf{C}_{k}^T(\mathbf{C}_{k}\mathbf{P}_k\mathbf{C}_{k}^T+\mathbf{\Sigma}_{\mathbf{w}_k})^{-T}\mathbf{C}_k\mathbf{P}_k^T, \label{f6}
\end{equation}
where $\mathbf{P}_k$ is the solution to the algebraic Riccati equation:
\begin{equation}
\label{k5}
\begin{aligned}
\mathbf{P}_k=&\mathbf{A}_k\mathbf{P}_k\mathbf{A}_k^T+\mathbf{\Sigma}_{\mathbf{v}_k}\\
&\quad-\mathbf{A}_k\mathbf{P}_k\mathbf{C}_{k}^T(\mathbf{C}_{k}\mathbf{P}_k\mathbf{C}_{k}^T+\mathbf{\Sigma}_{\mathbf{w}_k})^{-T}\mathbf{C}_k\mathbf{P}_k^T\mathbf{A}_k^T.
\end{aligned}
\end{equation}
Considering \eqref{R1} is nearly a tight lower bound of the LQR cost \cite{f12},
we thereby introduce the rate-cost function,
\begin{equation}
\label{e1}
\begin{aligned}
L(n_k,p_k)&\triangleq\frac{m |\det \mathbf{N}_k\mathbf{M}_{k}|^\frac{1}{m}} {2^{\frac{2}{m}(R(n_k,p_k)-\log_2|\det \mathbf{A}_k|)}-1}\\
&\quad\quad+\text{tr}\left( \mathbf{\Sigma}_{\mathbf{v}_k}\mathbf{S}_k\right)+\text{tr}(\mathbf{\Sigma}_k\mathbf{A}_k\mathbf{M}_k\mathbf{A}_k),
\end{aligned}
\end{equation}
which establishes a bridge between the limit LQR cost and the cycle rate. %This connection allows us to orchestrate communication resources for the closed-loop control, which is detailed in the next Section.

\section{Control-Oriented Optimization of the Mother-Daughter Downlink}
\label{section 3}
Next, we propose our control-oriented optimization scheme. Compared with those schemes addressed for general IoT scenarios, our formulated problem address the extreme conditions in the space environment. Specifically, given the limited communication resources carried by the space robots, we directly take the control performance as the optimization objective.
Additionally, concerning the varying environment in space, the data transmission within the $\mathbf{SC}^3$ loop is characterized in the FBL regime. Based on these two special concerns, our formulated problem thus takes the downlink block length and transmit power as the optimization variables to minimize the sum LQR cost of $K$ $\mathbf{SC^3}$ loops:
\begin{subequations}
\label{P1}
\begin{align}
\mbox{(P1)} \ \ \min_{\mathbf{n},\mathbf{p}} \ \ &\sum_{k = 1}^{K} L(n_k,p_k)  \label{P1a} \\
\mbox{s.t.}\quad
&R(n_k,p_k)\geqslant \log_2 |\det \mathbf{A}_k|, \ \forall k, \label{p1c}\\
&\frac{n_k}{B}\leqslant T_k, n_k\in\mathbb{N}^+,\ \forall k,\label{pn}\\
%		&\sum_{k=1}^{K} B_k\leqslant B_{\text{max}}, \\
%		&p_k\geqslant0, \ \forall k \\
&\sum_{k = 1}^{K} p_k \leqslant P_{\text{max}}, \label{p1d}\\
&p_k\geqslant 0, \forall k,
\end{align}
\end{subequations}
where $\mathbf{n}\triangleq\{n_1,...,n_K\}$ and $\mathbf{p}\triangleq\{p_1,...,p_K\}$. (P1) is a nonlinear mixed-integer problem. Directly solving it bears high complexity. First, the optimal block length is determined by $T_k$,
\begin{equation}
n^*_k=\lfloor BT_{k}\rfloor. \label{f3}
\end{equation}
This is because $L(n_k,p_k)$ is a decreasing function of $n_k$. It can be proven by the rule of the monotonicity of composite functions, as the rate-cost function decreases with the cycle rate and the cycle rate increases with the block length.  %To minimize the LQR
After determining the block length, the remaining power allocation problem is given by
\begin{subequations}
\label{P2}
\begin{align}
\mbox{(P2)} \ \ \min_{\mathbf{p}} \ \ &\sum_{k = 1}^{K} L(p_k)   \label{P2a} \\
\mbox{s.t.}\quad
& R(p_k)\geqslant \log_2 |\det \mathbf{A}_k|, \ \forall k, \label{p2c}\\
%	&\sum_{k=1}^{K} B_k\leqslant B_{\text{max}},\\
%			&p_k\geqslant0, \ \forall k \\
&\sum_{k = 1}^{K} p_k \leqslant P_{\text{max}}, \label{p2d}\\
&p_k\geqslant 0, \forall k,
\end{align}
\end{subequations}
where $R(p_k)\triangleq R(n_k^*,p_k)$ and $L(p_k)\triangleq L(n_k^*,p_k)$.
To solve this problem, we first analyze their monotonicity and convexity-concavity. For the cycle rate $R(p_k)$,
we analyze the property of the achievable rate expression, i.e., $r(p_k)\triangleq r(n_k^*,p_k)$, as given in  \eqref{ro}, which holds the same property as $R(p_k)$. For ease of expression,  we rewrite $r(p_k)$ as a function of the SNR, denoted as $\gamma_k\triangleq\frac{g_kp_k}{\sigma^2}$,

\begin{equation}
\label{f22}
r(\gamma)=\log_2(1+\gamma)-\eta\sqrt{V(\gamma)},
\end{equation}
where  $\eta\triangleq\sqrt{\frac{1}{n^*}}Q^{-1}(\epsilon)$ and the subscript $k$ is omitted for simplicity.
%In the literature, the properties of the achievable rate have been investigated in \cite{ref19}. But the authors investigate $r(\gamma)$ when $\gamma\geqslant 0$, which includes both $r(\gamma)<0$ and $r(\gamma)\geqslant0$.
Since the data rate cannot be negative in practice, we discuss the monotonicity and concavity-convexity in the region of  $r(\gamma)\geqslant0$, which is given by {\bf{Theorem 1}}.

\begin{figure*}
	\centering
	\includegraphics[width=5.8 in]{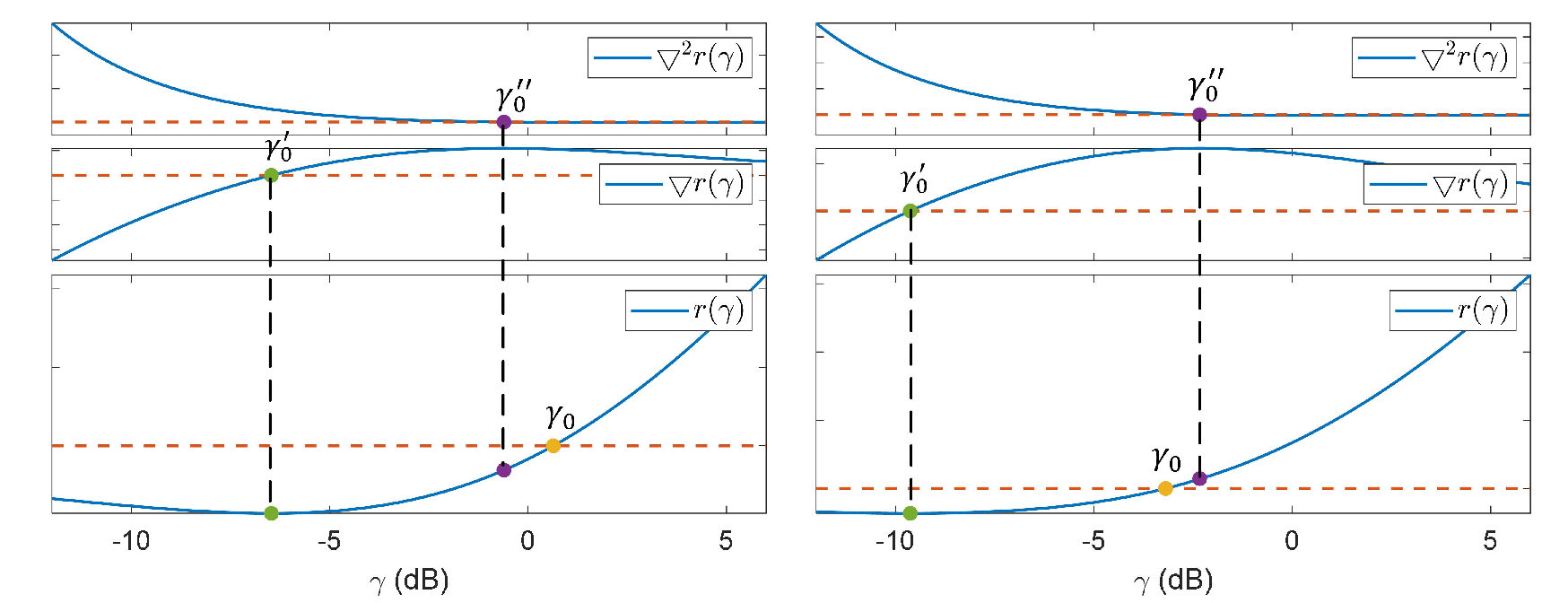}
	\caption{Curves of the achievable rate in the FBL regime, i.e.,  $r(\gamma)$, its first derivative, i.e., $\triangledown r(\gamma)$, and second derivative, i.e., $\triangledown^2 r(\gamma)$. Related parameters are set as: $\epsilon=10^{-6}$ and $n=30$ for the left figure, and $\epsilon=10^{-6}$ and $n=80$ for the right figure.}
	\label{figr}
\end{figure*}

\begin{theorem}
Denote the zero-crossing point of $r(\gamma)$ as $\gamma_0$ $(\gamma_0>0)$.
\begin{enumerate}
\item
$r(\gamma)$ is monotonically increasing with $\gamma$ in $[\gamma_0,+\infty)$.
\item
The convexity-concavity of $r(\gamma)$ is given by
\begin{equation}
	\left\{
	\begin{aligned}
		&r(\gamma) \  \text{is convex-concave in} \  [\gamma_0,+\infty), \quad \eta<\hat{\eta}, \\
		&r(\gamma) \ \text{is concave in} \ [\gamma_0,+\infty), \qquad\qquad\ \eta\geqslant \hat{\eta}.
	\end{aligned}
	\right.
\end{equation}
where  $\hat{\eta}\triangleq\frac{(1+\hat{\gamma})\ln(1+\hat{\gamma})}{\sqrt{\hat{\gamma}^2+2\hat{\gamma}}}$ and  $\hat{\gamma}$ is the zero-crossing point of the following equation
\begin{equation}
	\ln(1+\gamma)-\frac{(1+\gamma)(\gamma^2+2\gamma)^2}{3\gamma^2+6\gamma+1}=0.
\end{equation}
\end{enumerate}
\end{theorem}
\begin{IEEEproof}
See Appendix A.
\end{IEEEproof}
In Fig. \ref{figr}, we illustrate two examples of $r(\gamma)$ along with its first-order derivative $\triangledown r(\gamma)$  and second-order derivative $\triangledown^2 r(\gamma)$. The zero-crossing points of $\triangledown r(\gamma)$ and $\triangledown^2 r(\gamma)$ are denoted as $\gamma_0'$ and $\gamma_0''$, respectively. It is noted that
$\gamma_0'$ is the monotonicity turning point of $r(\gamma)$ and $\gamma_0''$ is the inflection point of $r(\gamma)$. The values of $\gamma_0'$ and $\gamma_0''$ compared with $\gamma_0$  determine the monotonicity and the concavity-convexity of $r(\gamma)$ in  $[\gamma_0, +\infty)$.
Observing both figures, we can see that there is $\gamma_0>\gamma_0'$ in both cases. This shows that $r(\gamma)$ is monotonically increasing in $[\gamma_0, +\infty)$ in  provided examples.
%In terms of the concavity-convexity, there is $\gamma_0>\gamma_0''$ in the left figure and $\gamma_0<\gamma_0''$ in the right figure. As a result, the two examples of $r(\gamma)$ are concave and convex-concave in $[\gamma_0, +\infty)$ respectively.
Moreover, $r(\gamma)$ demonstrates the concave nature in $[\gamma_0, +\infty)$ in the left example, as $\gamma_0$ exceeds $\gamma_0''$. Conversely, $r(\gamma)$ is convex-concave in  $[\gamma_0, +\infty)$ in the right example, as $\gamma_0$ is smaller than $\gamma_0''$.

Moving forward, based on the monotonicity of $r(\gamma)$ in $[\gamma_0, +\infty)$, the constraint \eqref{p2c} can be simplified into a threshold form
\begin{equation}
\label{thre}
p_k\geqslant p^{th}_k, \ \forall k,
\end{equation}
where $p^{th}_k$ is the solution to the following equation
\begin{equation}
\label{f18}
R(p_k)- \log_2 |\det \mathbf{A}_k|=0.
\end{equation}
This solution can be found by the dichotomy method. Thus, (P2) is simplified as
\begin{subequations}
\label{P3}
\begin{align}
\mbox{(P3)} \ \ \min_{\mathbf{p}} \ \ &\sum_{k = 1}^{K} L(p_k)   \label{P3a} \\
\mbox{s.t.}\quad
& p_k\geqslant p_k^{th}, \ \forall k, \label{p3c}\\
&\sum_{k = 1}^{K} p_k \leqslant P_{\text{max}}. \label{p3d}
\end{align}
\end{subequations}
%We omit the constraint $p\geqslant 0$ for that there must be $p_k^{th}>0$ when the system is unstable, i.e., ($\lambda_(\mathbf{A}_k)$)
In (P3), the stumbling block in solving the problem is the rate-cost function. It is a composite expression that includes the exponential, fractional, and logarithmic terms. Moving forward, we have {\bf{Theorem 2}}.
\begin{theorem}
The rate-cost function is a concave function with respect to the transmit power.
\end{theorem}
\begin{IEEEproof}
See Appendix B.
\end{IEEEproof}
Based on {\bf{Theorem 2}}, we can conclude that (P3) is a convex problem, which can be easily solved by the convex optimization tool. The corresponding algorithm is summarized in \textbf{Algorithm \ref{Tab1}}.

\section{Comparisons With Two Communication-Oriented Schemes}
\label{section 4}
In this section, we first investigate two communication-oriented schemes for the considered setting. They are the max-sum rate scheme and the max-min rate scheme. On this basis, we compare the proposed control-oriented scheme with these two schemes. Since the optimal block length is calculated using \eqref{f3},  we only focus on the power allocation in this section. %We first derive the closed-form solutions of these three schemes and subsequently analyze their power allocation principles.
%	In this section, we present an approximated close-form solution in the assure-to-be-stable condition. We also compare
\begin{algorithm}[t]
	\caption{Algorithm for Proposed Control-Oriented Communication Scheme } \small
	\label{Tab1}
	\begin{algorithmic}[1]
		\REQUIRE
		{
			The number of $\mathbf{SC}^3$ loops: $K$;\\
			Sensing-related parameters: $\mathbf{C}_k$ and $\mathbf{\Sigma}_{\mathbf{w}_k}$;\\
			Communication-related parameters: $B$, $P_{\max}$, $G_{\max}$, $\theta_{\text{3dB}}$, $\epsilon_k$, $d_k$, $f_c$, $\sigma^2$, $P_{Los}$, $SF_{1/2}$, and $CL$;\\
			Control-related parameters: $\mathbf{A}_k$, $\mathbf{B}_k$,  $\mathbf{Q}_k$, $\mathbf{R}_k$, $\mathbf{\Sigma}_{\mathbf{v}_k}$, and $m$;\\
			Cycle time: $T_k^0$ and the time for sensing, computing, and executing control actions: $t_k^s$, $t_{k}^{c}$, and $t^{a}_k$.\\
		}
		\STATE Calculate the  maximum allowable time for transmission, i.e.,  $T_k$, according to \eqref{rv1};
		\STATE  Calculate $\mathbf{S}_k$,  $\mathbf{M}_k$, $\mathbf{N}_k$, and $\mathbf{\Sigma}_k$ according to \eqref{Riccati}-\eqref{k5};
		\STATE Calculate the optimal block length $n_k^*$ according to \eqref{f3};
		\STATE Calculate the stable condition threshold $p_k^{th}$ according to \eqref{f18};
		%	\STATE Obtain $p_0$ according to \eqref{f18};
		\STATE  Obtain the optimal transmit power $p_k^{*}$ by solving (P3);
		\ENSURE	the optimal block length $n_k^*$ and transmit power $p_k^*$.
	\end{algorithmic}
\end{algorithm}
\subsection{ Communication-Oriented Power Allocation}
If we solely chase the optimum of communication, the data rate is a key metric that reflects the communication efficiency. The max-sum rate scheme and max-min rate scheme are two classical communication-oriented schemes. For the considered setting, the corresponding optimization problems are given by
%n\log_2(1+\frac{g_kp_k}{\sigma^2})-\sqrt{n}\log_2(e)Q(\epsilon_k),
\begin{subequations}
\begin{align}
\mbox{(PA)} \ \max_{\mathbf{p}}\ &\sum\limits_{k=1}^K R(p_k) \\
\quad \quad \mbox{s.t.}
&\ R(p_k)\geqslant\log_2|\text{det}\mathbf{A}_k|, \ \forall k, \label{f11} \\
& \sum_{k=1}^K p_k\leqslant P_{\max},\\
& p_k\geqslant 0, \forall k.
\end{align}
\end{subequations}
\begin{subequations}
\begin{align}
\mbox{(PB)} \ \max\limits_{\mathbf{p}}\min_k \ \  &R(p_k) \\
\quad \quad \quad \quad \mbox{s.t.}
\ &R(p_k)\geqslant\log_2|\text{det}\mathbf{A}_k|, \ \forall k, \label{f12} \\
&\sum_{k=1}^K p_k\leqslant P_{\max}\\
& p_k\geqslant 0, \forall k.
\end{align}
\end{subequations}
As stated in {\bf{Theorem 1}}, the cycle rate $R(p_k)$ is not necessarily concave with respect to the transmit power. This indicates that (PA) and (PB) are both non-convex. To address this obstacle, we employ Taylor expansion to linearize the nonconvex term of the cycle rate, i.e., $\sqrt{V(p_k)}$. The first-order Taylor expansion at the given point $\hat{p}_k$ is given by
\begin{equation}
\begin{aligned}
R(p_k|\hat{p}_k)&=(1-\epsilon_k)\big[ n_k^*\log_2(1+\frac{g_kp_k}{\sigma^2})-\\
&\sqrt{n_k^*}Q^{-1}(\epsilon_k)\big(\sqrt{V(\hat{p}_k)}+\bigtriangledown \sqrt{V(\hat{p}_k)}(p_k-\hat{p}_k)\big)\big],
\end{aligned}
\end{equation}
where
\begin{equation}
\bigtriangledown\sqrt{V(\hat{p}_k)}=\frac{\log_2(e)\frac{g_k}{\sigma^2}}{(1+\frac{g_k\hat{p}_k}{\sigma^2})^2\sqrt{(\frac{g_k\hat{p}_k}{\sigma^2})^2+2\frac{g_k\hat{p}_k}{\sigma^2}}}.
\end{equation}
On this basis, we can solve (PA) and (PB) in an iterative way. The problems at the $i$ th iteration are formulated as
\begin{subequations}
\begin{align}
\mbox{(PA-2)} \ \max_{\mathbf{p}}\ &\sum\limits_{k=1}^K R(p_k|p_k^{i-1}) \\
\quad \quad \mbox{s.t.}	&\ p_k\geqslant p_k^{th}, \ \forall k, \label{f111} \\
&\sum_{k=1}^K p_k\leqslant P_{\max}.
\end{align}
\end{subequations}
\begin{subequations}
\begin{align}
\mbox{(PB-2)} \ \max_{\mathbf{p}}\min_k \ \  &R(p_k|p_k^{i-1}), \\
\quad \quad \quad \quad \mbox{s.t.}	&\ p_k\geqslant p_k^{th}, \ \forall k,\label{f122}  \\
&\sum_{k=1}^K p_k\leqslant P_{\max},
\end{align}
\end{subequations}
where \eqref{f111} and \eqref{f122} are derived from the stable condition \eqref{f18} and $p_k^{i-1}$ is the solution obtained in the $(i-1)$ th iteration. The algorithm is assured to converge because of the monotonicity and boundedness of generated solutions. We summarize the corresponding algorithm in \textbf{Algorithm \ref{Tab2}}.
\begin{algorithm}[t]
\caption{Algorithm for the Max-Sum Rate and Max-Min Rate Schemes} \small
\label{Tab2}
\begin{algorithmic}[1]
\REQUIRE
{
	The number of $\mathbf{SC}^3$ loops: $K$;\\
	Sensing-related parameters: $\mathbf{C}_k$ and $\mathbf{\Sigma}_{\mathbf{w}_k}$;\\
	Communication-related parameters: $B$, $P_{\max}$, $G_{\max}$, $\theta_{\text{3dB}}$, $\epsilon_k$, $d_k$, $f_c$, $\sigma^2$, $P_{Los}$, $SF_{1/2}$, and $CL$;\\
	Control-related parameters: $\mathbf{A}_k$, $\mathbf{B}_k$, $\mathbf{Q}_k$, $\mathbf{R}_k$, $\mathbf{\Sigma}_{\mathbf{v}_k}$, and $m$;\\
	Cycle time: $T_k^0$ and the time for sensing, computing, and executing control actions: $t_k^s$, $t_{k}^{c}$, and $t^{a}_k$;\\
	Iteration termination threshold: $\delta$.
}
\STATE Calculate the  maximum allowable time for transmission, i.e.,  $T_k$,  according to \eqref{rv1};
\STATE  Calculate $\mathbf{S}_k$, $\mathbf{M}_k$, $\mathbf{N}_k$, and $\mathbf{\Sigma}_k$ according to \eqref{Riccati}-\eqref{k5};
\STATE Calculate the optimal block length $n_k^*$ according to \eqref{f3};
\STATE Calculate the stable condition threshold $p_k^{th}$ according to \eqref{f18};
\STATE \emph{Initialization}: $i=0$ and $p_k^0=p_k^{th}$;
\REPEAT
\STATE  $i=i+1$;
\STATE  Obtain $p_k^{(i)}$ by solving (PA-2) and (PB-2);
\UNTIL{$\frac{\abs{\sum\limits_{k=1}^KR(p_k^{(i)})-\sum\limits_{k=1}^KR(p_k^{(i-1)})}}{\sum\limits_{k=1}^KR(p_k^{(i-1)})}\leqslant \delta$ for (PA);  \\
	$\quad \quad \ \frac{\abs{\min\limits_{k}R(p_k^{(i)})-\min\limits_{k}R(p_k^{(i-1)})}}{\min\limits_{k}R(p_k^{(i-1)})}\leqslant \delta$ for (PB).}
\ENSURE	the optimal block length $n_k^*$ and transmit power $p_k^*$.
\end{algorithmic}
\end{algorithm}

%\begin{equation}
%	\sum_{k=1}^{K}R(p_k^i) \geqslant \sum_{k=1}^{K}R(p_k^i|p_k^{i-1}) \geqslant	\sum_{k=1}^{K}R(p_k|p_k^{i-1}) \geqslant \sum_{k=1}^{K}R(p_k^{i-1}|p_k^{i-1})=\sum_{k=1}^{K}R(p_k^{i-1})
%\end{equation}
\subsection{Approximate Closed-Form Solutions}
In this subsection, we derive the approximate closed-form solutions for three discussed schemes.  We assume that all the $\mathbf{SC}^3$ loops have the same maximum allowable transmission time, i.e., $T_k=T, \forall k$. According to \eqref{f3}, the optimal block length is calculated by $n=\lfloor BT\rfloor$. In addition, when the received SNR is greater than $10$ dB, i.e., $10\log_{10}(\frac{g_kp_k}{\sigma^2})\geqslant 10$ dB, the term $\sqrt{V(p_k)}$ in \eqref{r} is very close to 1, i.e., $\sqrt{V(p_k)}|_{SNR\geqslant 10 dB}\geqslant 0.9959$. Under this condition, the cycle rate expression can be approximated as follows \cite{f49}:
\begin{equation}
\label{f13}
R(p_k)\approx n\log_2(1+\frac{g_kp_k}{\sigma^2})-\sqrt{n}\log_2(e)Q^{-1}(\epsilon_k).
\end{equation}
Furthermore, if the system further works in the assure-to-be-stable region, that the cycle rate is much larger than the intrinsic entropy, i.e.,   $R(p_k)\gg\log_2|\det\mathbf{A}_k|$, we can omit the term of minus one in the numerator in the rate-cost function and obtain
the following approximation:
\begin{equation}
\label{e2}
\begin{aligned}
L(p_k)&=\frac{m |\det \mathbf{N}_k\mathbf{M}_{k}|^\frac{1}{m}} {2^{\frac{2}{m}(R(p_k)-\log_2|\det \mathbf{A}_k|)}-1}\\
&\quad\quad+\text{tr}\left( \mathbf{\Sigma}_{\mathbf{v}_k}\mathbf{S}_k\right)+\text{tr}(\mathbf{\Sigma}_k\mathbf{A}_k\mathbf{M}_k\mathbf{A}_k)\\
&\approx \frac{m |\det \mathbf{N}_k\mathbf{M}_{k}|^\frac{1}{m}} {2^{\frac{2}{m}(R(p_k)-\log_2|\det \mathbf{A}_k|)}}\\
&\quad\quad+\text{tr}\left( \mathbf{\Sigma}_{\mathbf{v}_k}\mathbf{S}_k\right)+\text{tr}(\mathbf{\Sigma}_k\mathbf{A}_k\mathbf{M}_k\mathbf{A}_k),
\end{aligned}
\end{equation}
By combining \eqref{f13} and \eqref{e2}, we can get the following approximation of the rate-cost expression:
\begin{equation}
	\label{j5}
\begin{aligned}
L(p_k)&\approx\frac{m |\det \mathbf{N}_k\mathbf{M}_{k}|^\frac{1}{m}|\text{det}\mathbf{A}_k|^\frac{2}{m}e^{\frac{2\sqrt{n}}{m}Q^{-1}(\epsilon_k)}}{(1+\frac{gp_k}{\sigma^2})^\frac{2n}{m}}\\
&\quad\quad+\text{tr}\left( \mathbf{\Sigma}_{\mathbf{v}_k}\mathbf{S}_k\right)+\text{tr}(\mathbf{\Sigma}_k\mathbf{A}_k\mathbf{M}_k\mathbf{A}_k).\\
\end{aligned}
\end{equation}
Based on above approximation processes, we can see that \eqref{j5} closely aligns to the original expression when the system is in the assure-to-be-stable region. However,  this approximation would lose its efficiency  when the working point is out of this region.  Since the following introduced closed-form solution is derived from the approximation of the rate-cost function, it works well in the assure-to-be-stable region where the accuracy of these approximations are maintained. Correspondingly, the precision of the closed-form solution diminishes in the region approaching the critical stability, where the approximations are no longer accurate.
By  testing the Slater condition, it is easy to find that the strong duality holds for (P3). Solving its dual problem yields the same optimal solution. The dual problem is given by
\begin{subequations}
\begin{align}
\mbox{(P4)} \quad \max_{\lambda}\min_{\mathbf{p}} \quad& \sum_{k=1}^K L(p_k)+\lambda(\sum_{k = 1}^{K} p_k-P_{\text{max}})\\
\mbox{s.t.}\quad
%		R(n_k^*,p_k) >\log_2|\det\mathbf{A}_k|\label{nec},  \quad \forall k\\
&\lambda \geqslant 0,
\end{align}
\end{subequations}
where $\lambda$ is the Lagrangian multiplier. The stable condition is omitted because it is satisfied in the assure-to-be-stable region. The  Karush-Kuhn-Tucker (KKT) conditions of (P4) are given by
\begin{subequations}
\begin{numcases}{}
\frac{\partial L(p_k)}{\partial p_k}+\lambda\bigg|_{p_k=p_k^*}=0, \forall k, \label{cd1}\\
\sum_{k = 1}^{K} p_k-P_{\text{max}}\bigg|_{p_k=p_k^*} = 0, \label{cd2}\\
\lambda\geqslant0,
\end{numcases}
\end{subequations}
where the equality of \eqref{cd2} comes from the monotonicity of $L(p_k)$. Using the approximation of $L(p_k)$ in \eqref{e2}, we calculate $\frac{\partial L(p_k)}{\partial p_k}$. Then,  \eqref{cd1} can be reorganized into:
\begin{equation}
\label{p_op}
p_k^*(\lambda)=(\frac{\sigma^2}{g_k})^\frac{2n}{2n+m}(\frac{2nG_k}{\lambda})^{\frac{m}{2n+m}}e^\frac{2\sqrt{n}Q^{-1}(\epsilon_k)}{2n+m}-\frac{\sigma^2}{g_k},
\end{equation}
where  $G_k\triangleq |\det \mathbf{N}_k\mathbf{M}_{k}|^\frac{1}{m}|\det\mathbf{A}_k|^\frac{2}{m}$,
which represents the sensing-and-control-related parameter.
By substituting \eqref{p_op} into \eqref{cd2}, we can further obtain that
\begin{equation}
(\frac{1}{\lambda})^{\frac{m}{2n+m}}=\frac{P_{\max}+\sum\limits_{i=1}^K\frac{\sigma^2}{g_i}}{\sum\limits_{i=1}^K(\frac{\sigma^2}{g_i})^{\frac{2n}{2n+m}}(2nG_i)^{\frac{m}{2n+m}}e^\frac{2\sqrt{n}Q^{-1}(\epsilon_i)}{2n+m}} \label{cd4}.
\end{equation}
Combining  \eqref{p_op} and \eqref{cd4}, the closed-form solution to the proposed scheme is derived,
\begin{equation}
\label{op1}
p_k^*=(P_{\max}+\sum_{i=1}^K\frac{\sigma^2}{g_i})\frac{(\frac{\sigma^2}{g_k})^\frac{2n}{2n+m}e^\frac{2\sqrt{n}Q^{-1}(\epsilon_k)}{2n+m}G_k^\frac{m}{2n+m}}{\sum\limits_{i=1}^K(\frac{\sigma^2}{g_i})^\frac{2n}{2n+m}e^\frac{2\sqrt{n}Q^{-1}(\epsilon_i)}{2n+m}G_i^\frac{m}{2n+m}}-\frac{\sigma^2}{g_k}.
\end{equation}
In addition, when the cycle-time differences of different $\mathbf{SC}^3$ loops are not significant, \eqref{op1} can be applied to the more general case in which different $\mathbf{SC}^3$ loops have different cycle time by substituting $n$ with $n_k^*$.

Moving forward, we derive the closed-form solutions for two communication-oriented schemes. %The stable condition \eqref{f11} and \eqref{f12} is transformed into a threshold constraint as shown in \eqref{thre}.
Using the high-SNR approximation in \eqref{f13}, (PA) and (PB) become two convex problems. As for (PA), if without the stable condition \eqref{f11},  the optimal solution is given by the classical Water-Filling method \cite[Ch. 9.4]{f51}
\begin{equation}
\label{op2}
p_k^*=\big[\frac{1}{\lambda}-\frac{\sigma^2}{g_k}\big]^{+},
\end{equation}
where $[x]^+=\max\{x,0\}$, $\lambda$ is chosen to satisfy $\sum\limits_{k=1}^Kp_k=P_{\max}$. On this basis, it is easy to derive the closed-form solution to (PA) by taking \eqref{f11} into account:
\begin{equation}
p_k^*=\max\bigg(\big[\frac{1}{\lambda}-\frac{\sigma^2}{g_k}\big]^{+}, p_k^{th}\bigg),
\end{equation}
where $\lambda$ is chosen to satisfy $\sum\limits_{k=1}^Kp_k=P_{\max}$.
%If $P_{\max}$ is big and $p_k\geqslant p_k^{th}$ is satisfied, the closed-form solution can be explicitly expressed as
%\begin{equation}
%	p_k^*=\frac{P_{\max}+\sum\limits_{i=1}^K\frac{\sigma^2}{g_i}}{K}-\frac{\sigma^2}{g_k}.
%\end{equation}

As for the max-min rate scheme, if there is no stable condition \eqref{f12}, the optimal solution to (PB) satisfies
\begin{equation}
R(p^*_1)=R(p^*_2)\cdots=R(p^*_K)=\lambda, \label{f14}
\end{equation}
where $\lambda$ is an auxiliary variable. This is because if \eqref{f14} does not hold, the minimal cycle rate among the $K$ $\mathbf{SC}^3$ loops could be further improved by reallocating the transmit power from other loops to the loop with the minimum cycle rate. Based on \eqref{f14}, we could derive that
\begin{equation}
\label{f15}
p_k^*=2^{\frac{\lambda}{n}}(\frac{\sigma^2}{g_k})e^\frac{Q^{-1}(\epsilon_k)}{\sqrt{n}}-\frac{\sigma^2}{g_k}.
\end{equation}
Then, substituting \eqref{f15} into $\sum\limits_{k=1}^Kp_k^*=P_{\max}$, we can obtain that
\begin{equation}
\label{f21}
2^{\frac{\lambda}{n}}=\frac{P_{\max}+\sum\limits_{i=1}^K\frac{\sigma^2}{g_i}}{\sum\limits_{i=1}^K(\frac{\sigma^2}{g_i})e^\frac{Q^{-1}(\epsilon_i)}{\sqrt{n}}}.
\end{equation}
We further substitute \eqref{f21} into \eqref{f15}. The closed-form solution is given by
\begin{equation}
\label{op3}
p_k^*=(P_{\max}+\sum\limits_{i=1}^K\frac{\sigma^2}{g_i})\frac{(\frac{\sigma^2}{g_k})e^\frac{Q^{-1}(\epsilon_k)}{\sqrt{n}}}{\sum\limits_{i=1}^K(\frac{\sigma^2}{g_i})e^\frac{Q^{-1}(\epsilon_i)}{\sqrt{n}}}-\frac{\sigma^2}{g_k}.
\end{equation}
On this basis, it is easy to obtain the closed-form solution to (PB) by taking the stable condition into account,
\begin{equation}
p_k^*=\max\bigg[(\lambda+\sum\limits_{i=1}^K\frac{\sigma^2}{g_i})\frac{(\frac{\sigma^2}{g_k})e^\frac{Q^{-1}(\epsilon_k)}{\sqrt{n}}}{\sum\limits_{i=1}^K(\frac{\sigma^2}{g_i})e^\frac{Q^{-1}(\epsilon_i)}{\sqrt{n}}}-\frac{\sigma^2}{g_k}, \ p_k^{th}\bigg],
\end{equation}
where $\lambda$ is chosen to satisfy $\sum\limits_{k=1}^Kp_k^*=P_{\max}$. In addition, if $P_{\max}$ is large enough  such that $p_k\geqslant p_k^{th}$ is satisfied, the optimal solution is exactly \eqref{op3}.

\subsection{Analysis of Power Allocation Principles}
Comparing the closed-form solutions of the control-oriented scheme, the max-sum rate scheme and the max-min rate scheme, we have following observations:
\begin{enumerate}
\item The  proposed control-oriented scheme provides a way to account for sensing and control factors in the communication design. According to \eqref{op1}, the $\mathbf{SC^3}$ loops with less accurate sensing and more instability (corresponding to a larger $G_k$) are assigned  more transmit power. In contrast, the two communication-oriented schemes overlook these crucial factors as they solely focus on the communication process.
\item In terms of channel conditions, the proposed scheme and the max-min rate scheme exhibit a similar power allocation pattern. They allocate more power to the $\mathbf{SC^3}$ loops with poorer channel conditions. Conversely, the max-sum rate scheme allocates more power to the $\mathbf{SC^3}$ loops with better channel conditions.
\end{enumerate}
Inspired by the similar allocation principle between the control-oriented scheme and the max-min rate scheme in terms of channel conditions, we further find their equivalence for time-insensitive tasks, which is concluded by the  following {\bf{Theorem}}.
\begin{theorem}
When $n\gg m$, the following modified max-min rate scheme yields the same power allocation solution as the control-oriented scheme
\begin{subequations}
\begin{align}
	\label{app_control}
	\mbox{(PC)} \ \max\limits_{\mathbf{p}}\min\limits_{k}\ &R(p_k)-\frac{m}{2}\log_2(G_k) \\
	\quad \quad \text{s.t.}	\ &R(p_k)\geqslant\log_2|\det\mathbf{A}_k|, \ \forall k,  \\
	&\sum_{k=1}^K p_k\leqslant P_{\max},\\
	&p_k\geqslant 0, \forall k.
\end{align}
\end{subequations}
In addition, when the optimal block length, i.e., $n$, goes to infinity, the max-min rate scheme yields the same solution as the proposed scheme.
\end{theorem}
%If $n$ goes to the infinity, the control-oriented scheme is equivalent to the max-min rate scheme.

\begin{IEEEproof}
When $n\gg m$, we have that
$\lim\limits_{n\gg m}2n+m=2n$.
Then, the closed-form solution of the proposed scheme \eqref{op1} has the following approximation:
\begin{small}
\begin{equation}
	\label{c1}
	\begin{aligned}
		\lim_{n\gg m}\bigg[	(&P_{\max}+\sum_{i=1}^K\frac{\sigma^2}{g_i})\frac{(\frac{\sigma^2}{g_k})^\frac{2n}{2n+m}e^\frac{2\sqrt{n}Q^{-1}(\epsilon_k)}{2n+m}G_k^\frac{m}{2n+m}}{\sum\limits_{i=1}^K(\frac{\sigma^2}{g_i})^\frac{2n}{2n+m}e^\frac{2\sqrt{n}Q^{-1}(\epsilon_i)}{2n+m}G_i^\frac{m}{2n+m}}-\frac{\sigma^2}{g_k}\bigg],  \\	=	(&P_{\max}+\sum_{i=1}^K\frac{\sigma^2}{g_i})\frac{(\frac{\sigma^2}{g_k})e^\frac{Q^{-1}(\epsilon_k)}{\sqrt{n}}G_k^\frac{m}{2n}}{\sum\limits_{i=1}^K(\frac{\sigma^2}{g_i})e^\frac{Q^{-1}(\epsilon_i)}{\sqrt{n}}G_i^\frac{m}{2n}}-\frac{\sigma^2}{g_k}.
	\end{aligned}
\end{equation}
\end{small}
By comparing \eqref{c1} with the closed-form solution of the max-min rate scheme \eqref{op3}, it is easy to find that \eqref{c1} is the closed-form solution to the modified max-min problem (PC).
In addition, when the block length $n$ goes to infinity, the following approximations can be made:
\begin{equation}
\begin{aligned}
	&\lim\limits_{n\rightarrow +\infty}\frac{2n}{2n+m}=1,\\
	&\lim\limits_{n\rightarrow +\infty}\frac{m}{2n+m}=0,\\
	&\lim\limits_{n\rightarrow +\infty}\frac{2\sqrt{n}}{2n+m}=\frac{1}{\sqrt{n}}.
\end{aligned}
\end{equation}
Then, the closed-form solution to the proposed scheme \eqref{op1} can be further simplified into
\begin{small}
\begin{equation}
\begin{aligned}
	\lim_{n\rightarrow +\infty}\bigg[	&(P_{\max}+\sum_{i=1}^K\frac{\sigma^2}{g_i})\frac{(\frac{\sigma^2}{g_k})^\frac{2n}{2n+m}e^\frac{2\sqrt{n}Q^{-1}(\epsilon_k)}{2n+m}G_k^\frac{m}{2n+m}}{\sum\limits_{i=1}^K(\frac{\sigma^2}{g_i})^\frac{2n}{2n+m}e^\frac{2\sqrt{n}Q^{-1}(\epsilon_i)}{2n+m}G_i^\frac{m}{2n+m}}-\frac{\sigma^2}{g_k}\bigg], \\
	=(&P_{\max}+\sum_{i=1}^K\frac{\sigma^2}{g_i})\frac{(\frac{\sigma^2}{g_k})e^\frac{Q^{-1}(\epsilon_k)}{\sqrt{n}}}{\sum\limits_{i=1}^K(\frac{\sigma^2}{g_i})e^\frac{Q^{-1}(\epsilon_i)}{\sqrt{n}}}-\frac{\sigma^2}{g_k}, \label{k4}
\end{aligned}
\end{equation}
\end{small}
which is the same as the solution to the max-min rate scheme, i.e., \eqref{op3}.
\end{IEEEproof}
{\bf{Theorem 3}} reveals the fairness-minded nature of the proposed scheme. In fact, to avoid the ``short-board effect" that drags down the system efficiency, the proposed scheme guarantees the basic performance of all loops.  In addition, the equivalence  between the proposed scheme and the max-min rate scheme indicates that as the cycle time increases, the interdependence between communication and control weakens. Under this condition, the max-min rate scheme yields the near-optimal solution for the closed-loop control. Therefore, we can conclude that the proposed scheme is necessary for time-sensitive control tasks, while the max-min rate scheme is a good alternative for time-insensitive cases.

\section{Simulation Results and Discussion}
\label{section 5}
In the simulation, we assume that there are $K=4$ $\mathbf{SC^3}$ loops.
For sensing-related parameters, we set $\mathbf{C}_k=\mathbf{I}_m$ and $\mathbf{\Sigma}_{\mathbf{w}_k}=\mathbf{0}_m$. For control-related parameters, we set $m=100$, $\mathbf{Q}_k=\mathbf{I}_m$, $\mathbf{R}_k=\mathbf{0}_m$, and $\mathbf{\Sigma}_{\mathbf{v}_k}=0.01*\mathbf{I_m}$ \cite{ref21}. By setting $\mathbf{R}_{k}=\mathbf{0}_m$, the LQR cost quantifies state deviations, which directly reflects the control effectiveness of the $\mathbf{SC}^3$ loop. This accords to our focus of the control effectiveness rather than the control cost.
In the case of $\mathbf{R}_{k}\neq \mathbf{0}_m$, the LQR cost changes accordingly per \eqref{Riccati}. Corresponding simulation results will give the same conclusions as in the case of $\mathbf{R}_{k}=\mathbf{0}_m$, and we do not further discuss this case.  The intrinsic entropy is given by  $\log_2|\det\mathbf{A}_k|=\{48,36,24,12\}$ for $k=1\sim4$.  The control cycle time is $T_k^0=50$ ms and the time used for sensing, computing and control is $40$ ms. In this way,  the time allowed for communication is 10 ms, which accords to the closed-loop control requirement in space exploration  \cite{f52}. As for communication-related parameters,
each $\mathbf{SC^3}$ loop is allocated with a narrow frequency band of $B=15$ kHz. The carrier frequency is $f_c=2$ GHz, the variance of the channel noise is $\sigma^2=-110$ dBm \cite{ref21},  and the transmission error probability is  $\epsilon_k=10^{-6}$. The maximal antenna gain is $G_{\max}=23$ dBi and $\theta_{\text{3dB}}=30^{\circ}$. Given that there are no high buildings on the explored planet,
we use the rural scenario to model the near-planet environment. The channel-related parameters are given by   $P_{Los}=91.9\%$,  $SF_1=1.14$ dB, $SF_2=8.78$ dB and $CL=18.42$ dB \cite{ff8}.
According to the height of the Mars orbit, we set the height of the mother spacecraft to 3000 km and four probes are located $[1000, 2000, 3000, 4000]$ km away from the horizontal projection point of the spacecraft.  In the present figures, the dotted line represents the infinite LQR cost and the system is unstable in such cases.

%		\centering
%		\includegraphics[width=2.8 in]{fig3.eps}
%		\caption{The limit control performance with different maximal transmit power, varying with the bit error. }
%		\label{f3}
%	\end{figure}

\begin{figure}
	\centering
	\includegraphics[width=0.9\linewidth]{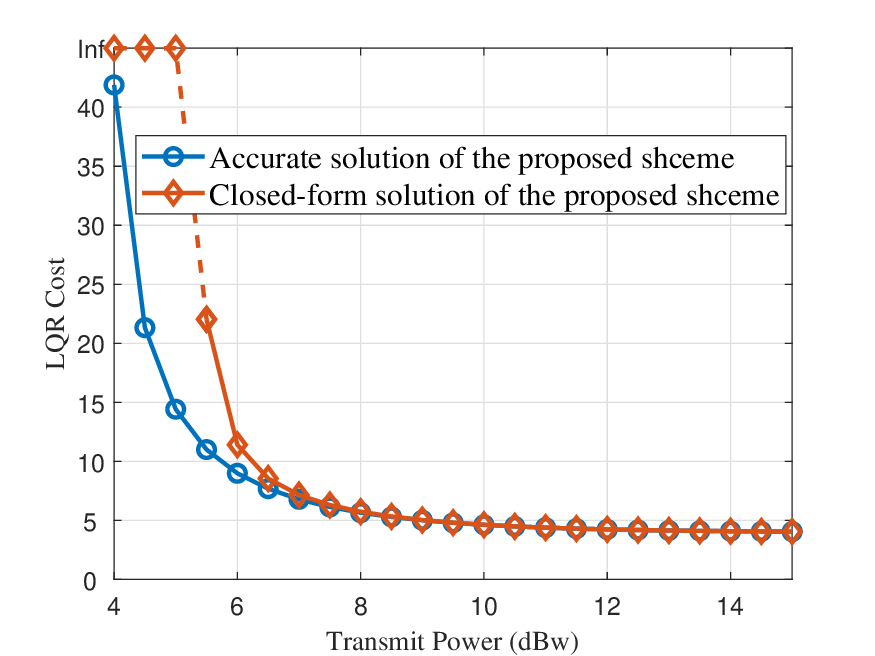}
	\caption{Comparisons of the LQR cost under the accurate solution and the proposed closed-form solution.}
	\label{ad11}
\end{figure}
In Fig. \ref{ad11}, we discuss the applicability of the proposed approximate closed-form solution \eqref{op1}. In this simulation, the minimal transmit power to stabilize the control system is 3.49 dBw. It can be seen that when $P_{\text{sum}}< 8 \text{dBw}$, the LQR cost under the closed-form solution is obvious higher than that under the optimal solution. This is because the approximations of the cycle rate \eqref{f13} and the rate-cost function \eqref{e2} deviate from their original expressions. In contrast, when $P_{\text{sum}}\geqslant 8 \text{dBw}$, the LQR cost under the  closed-form solution is nearly the same  as that under the optimal solution. This verifies the applicability of the closed-form solution in the assure-to-be-stable region.

\begin{figure}
	\centering
	\includegraphics[width=0.9\linewidth]{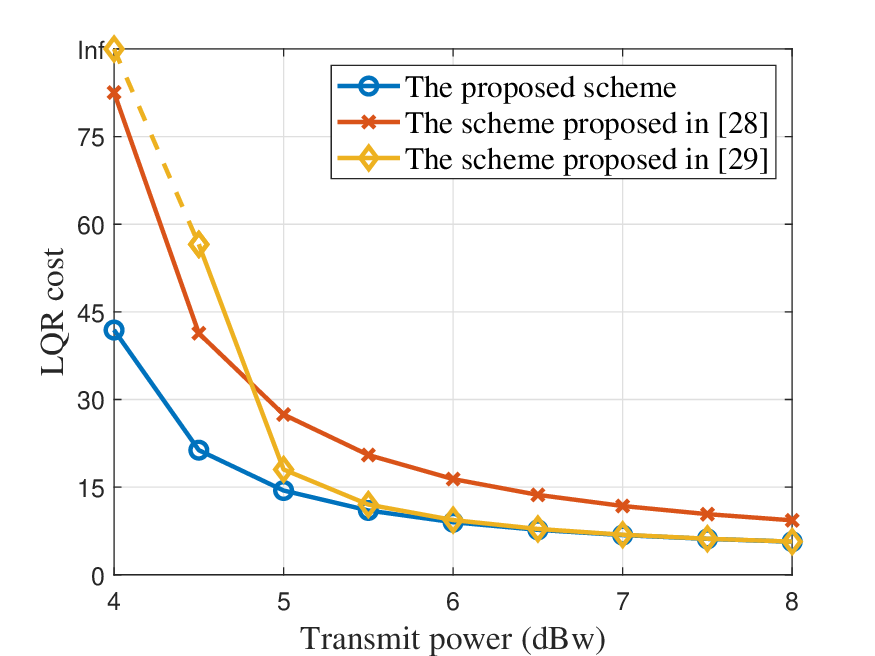}
	\caption{Comparisons of the LQR cost under three control-oriented  schemes. }
	\label{f1}
\end{figure}

In  Fig. \ref{f1}, we compare the proposed scheme with two control-aware schemes for general IoT scenarios \cite{ref11}\cite{ref21}.
In \cite{ref11}, the authors paid particular attention to the low-latency communication but they did not prioritize the task performance in the optimization design, using a hybrid metric named the energy-to-control efficiency. We thereby use  $\frac{P_{\max}-\sum\limits_{k=1}^K{p_k}}{\sum\limits_{k=1}^KL(n_k,p_k)}$ as its objective for a fair comparison. In contrast, the scheme proposed in \cite{ref21} directly optimized control LQR cost, but it overlooked the cycle-time constraint, applying the Shannon capacity to calculate the cycle rate. Since these two schemes were either not completely control oriented or time sensitive, the control performance of corresponding ``mother-daughter system" is inferior compared with the proposed scheme, making them less effective in the space setting.
It can be seen that the scheme proposed in \cite{ref21} performs the worst when $P_{\max}\leqslant 4.5$ dBw.
This is  attributed to its unawareness of the rate losses in the FBL regime. Some loops that should have been allocated with more power are under allocated. This leads to the system instability in the low-power region. As the power increases, the LQR cost under the scheme in \cite{ref11} becomes the highest. This is due to its emphasis on the energy efficiency. The control performance has to make a compromise to save power. These results validate  the effectiveness of the proposed scheme in the delay-sensitive  and resource-constrained space setting.

%{\color{blue} In short, since these two schemes either did not  directly optimize the control metric \cite{ref11} or did not model the data transmissions in the FBL regime \cite{ref21}, they can failed to provide the same control performance as the proposed scheme.}

\begin{figure}
\centering
\includegraphics[width=0.9\linewidth]{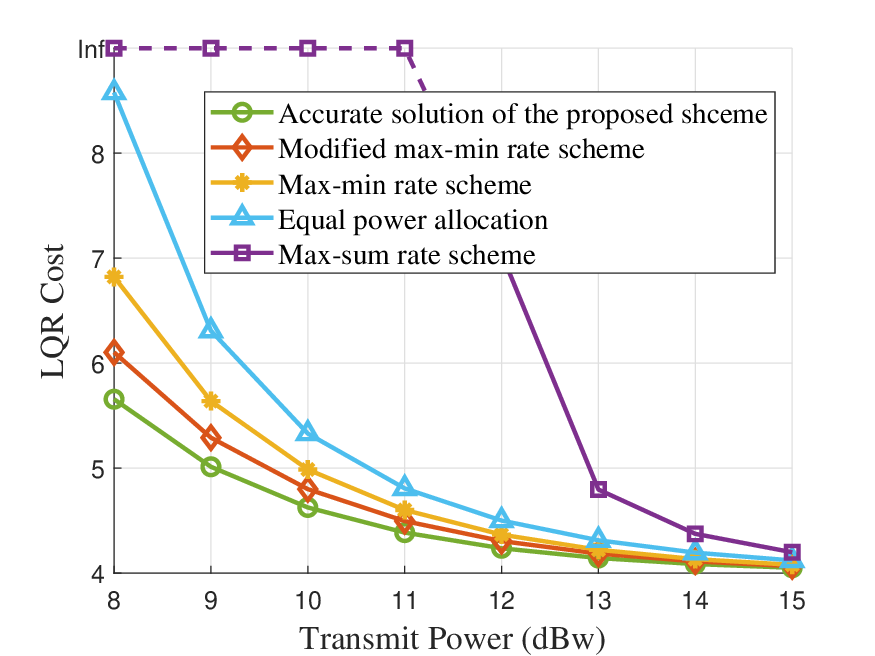}
\caption{Comparisons of the proposed scheme, the modified max-min rate scheme, the max-min rate scheme, the max-sum rate scheme and equal power allocation. }
\label{f2}
\end{figure}

In Fig. \ref{f2}, we compare the proposed scheme with three communication-oriented schemes: the max-sum rate scheme, the max-min rate scheme, and the classical equal power allocation. The curves under the approximate closed-form solution of the proposed scheme \eqref{op1} and the modified max-min rate scheme (PC) are also shown. %It can be seen that the control performance using the approximate closed-form solution \eqref{op1} is nearly the same as that using the accurate solution.
The performance under the modified max-min rate scheme shows a small gap from the optimal one. These results confirm their excellent approximations to the optimal solution. In addition, we can also see that the max-min rate scheme outperforms  equal power allocation. Both of them further outperform the max-sum rate scheme. This is because of their different power allocation principles
in terms of the channel condition. The max-min rate scheme holds a similar allocation principle as the proposed scheme, while the max-sum rate scheme follows an opposite allocation principle.

\begin{figure}
\centering
\includegraphics[width=0.9\linewidth]{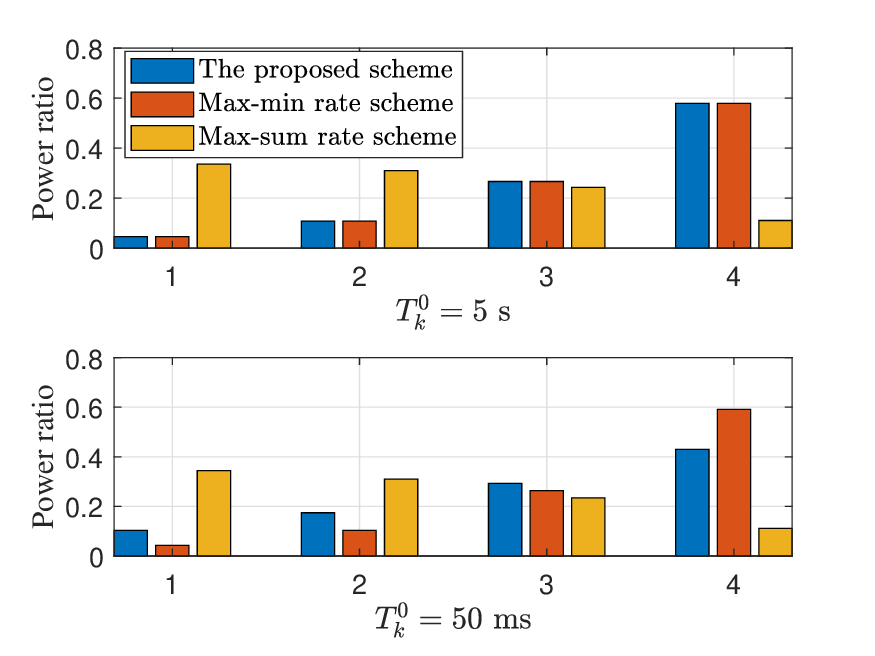}
\caption{The power allocation results under the proposed scheme, the max-sum rate scheme and the max-min rate scheme. }
\label{f4}
\end{figure}

In Fig. \ref{f4}, we present the power allocation results of the proposed scheme, the max-sum rate scheme, and the max-min rate scheme. The maximal transmit power is set as $P_{\max}=10$ dBw. It is evident that in both subfigures, the power ratio increases from loops one to four under the proposed scheme and the max-min rate scheme, while it decreases under the max-sum rate scheme.  This verifies their power allocation principles in terms of channel conditions, recalling $d_k=[1000,2000,3000,4000] \ \text{km}$.
In addition, when the cycle time is $50$ ms, the proposed scheme allocates more power to $\mathbf{SC^3}$ loops $1\sim3$ compared with the max-min rate scheme.   This is because we set $\log_2|\det\mathbf{A}_k|=\{48,36,24,12\}$ for $k=1\sim4$. This adjustment is made to compensate for the intrinsic entropy difference. In the top subfigure, which shows the case of $T_k^0=5$ s, the power allocation results of the proposed scheme and the max-min rate scheme are nearly the same.
In this situation, the intrinsic entropy difference has little impact on the communication power allocation. This verifies our conclusions drawn in {\bf{Theorem 3}}.

\begin{figure}
\centering
\includegraphics[width=0.9\linewidth]{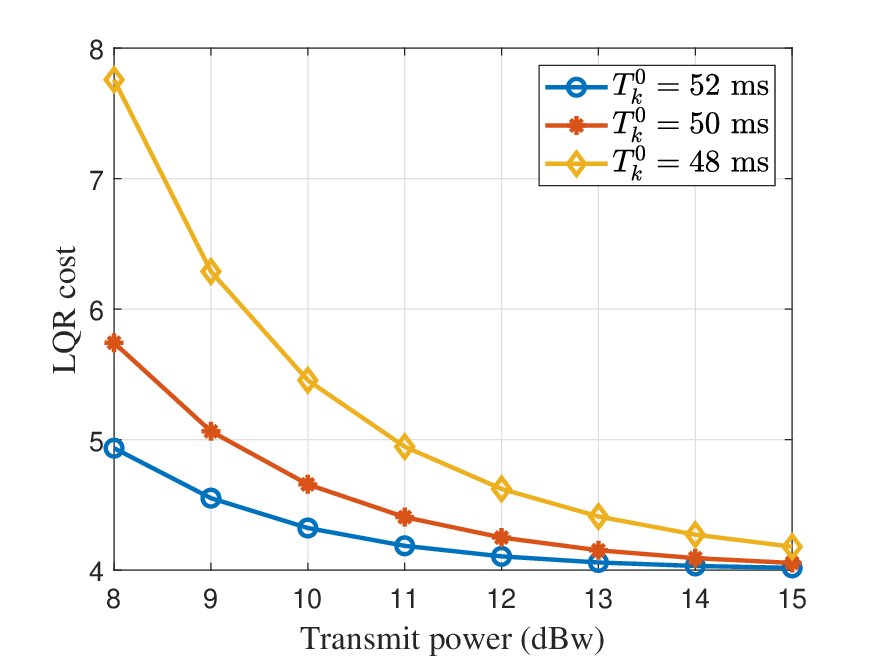}
\caption{Comparisons of the LQR cost under different cycle time}
\label{fig5}
\end{figure}

Next, we discuss the impacts of the cycle time  on the control performance.
The results are presented in Fig. \ref{fig5}. It can be seen that the cycle time impacts the control performance in the low-power region. As the power increases, different curves converge to the same minimal value. This is because in the low--power region, the cycle rate is very limited. Different cycle time brings with different cycle rates and different accuracy of the commands, which, consequently, leads to different control performances. While in the high-power region, all loops have sufficient cycle rates to deliver commands accurately. The control performances all go to the same optimum, making the cycle time difference less influential in this region.

\begin{figure}
\centering
\includegraphics[width=0.9\linewidth]{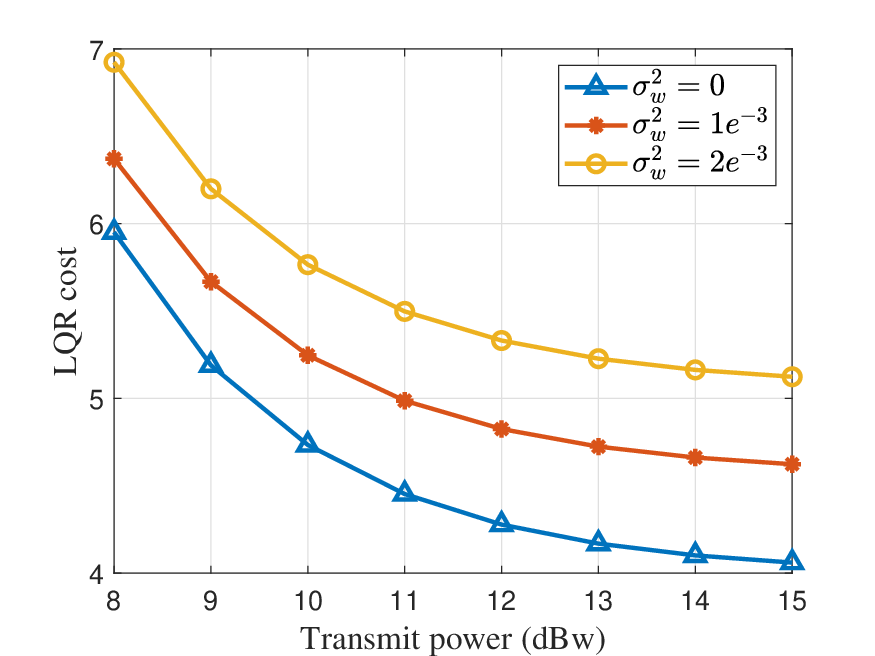}
\caption{Comparisons of the LQR cost under different sensing noise}
\label{fig6}
\end{figure}

In Fig. \ref{fig6},  we present the impacts of the sensing noise on the control performance. In this simulation, we set the sensing noise covariance as $\sigma^2_{w}=[0, 1e^{-3},2e^{-3}]$.
Different from the impact of cycle time, the LQR costs under different sensing noises would not converge to the same minimal value as the power increases.
This is because the inaccurate sensing would lead the control command deviating from the optimal one. This deviation cannot be  eliminated regardless of how excellent the communication is. This can also be explained by the rate-cost function \eqref{e1}.
The function's last term, i.e., $\text{tr}(\mathbf{\Sigma}_k\mathbf{A}_k\mathbf{M}_k\mathbf{A}_k)$, is determined by the sensing noise and control parameters and is irrelevant of the communication quality.
The larger value of $\sigma_{w}^2$ leads to the larger value of $\text{tr}(\mathbf{\Sigma}_k\mathbf{A}_k\mathbf{M}_k\mathbf{A}_k)$,
which can not be minimized by good communication.
In fact, in the flat region of these curves, the value of the limit LQR cost is determined by the last two terms of the rate-cost function. In this situation, the  bottleneck of the $\mathbf{SC}^3$ loop is sensing but not communication.

\section{Conclusions}
\label{section 6}
In this paper, we have investigated a control-oriented communication scheme for a space cooperation system named the ``mother-daughter system". The proposed scheme have optimized the block length and transmit power to minimize the sum LQR cost of the $\mathbf{SC^3}$ loops. The mother-daughter downlink has been modeled in the FBL regime. We have solved the nonlinear mixed-integer problem by proving the monotonicity and convexity of the achievable rate expression and the rate-cost function. The approximate closed-form solutions for the control-oriented scheme and two communication-oriented schemes: the max-sum rate scheme and the max-min rate scheme, have been derived.  Based on these closed-form solutions, we have revealed the equivalence between the proposed control-oriented scheme and the max-min rate scheme for time-insensitive control tasks. Simulation results have verified the effectiveness of the control-oriented communication scheme in improving the control task performance.

\appendices

\section{Proof of \bf{Theorem 1}}
To prove the monotonicity and concavity-convexity of the achievable rate expression in the FBL regime, i.e., $r(\gamma)$, as shown in \eqref{f22}, we first calculate its first-order and second-order derivatives
\begin{equation}
\triangledown r(\gamma)=\frac{\log_2(e)}{\gamma+1}\big(1-\frac{\eta}{(\gamma+1)\sqrt{\gamma^2+2\gamma}}\big),
\end{equation}
\begin{equation}
\label{57}
\begin{aligned}
&\triangledown^2 r(\gamma)=\frac{\log_2(e)}{(\gamma+1)^2}\big(\frac{2\eta}{(\gamma+1)(\gamma^2+2\gamma)^\frac{1}{2}}+\frac{\eta(\gamma+1)}{(\gamma^2+2\gamma)^\frac{3}{2}}-1\big).
\end{aligned}
\end{equation}
On this basis, it can be proven that $\triangledown r(\gamma)$ is thus first negative and then positive in $[0,+\infty)$, and we denote the zero-crossing point as $\gamma_0'$.
$r(\gamma)$ is monotonically decreasing in $[0, \gamma_0')$ and monotonically increasing in $[\gamma_0',+\infty)$.
Since $r(\gamma)\big|_{\gamma=0}=0$, $r(\gamma)$ is negative in $[0, \gamma_0')$. Therefore, the zero-crossing point of $r(\gamma)$, i.e., $\gamma_0$, is greater than $\gamma_0'$.  $r(\gamma)$ is  monotonically increasing in $[\gamma_0,+\infty)$.

Similarly, it can be proven that $\triangledown^2 r(\gamma)$ is first positive and then negative when $\gamma\in[0,+\infty)$. Denote its zero-crossing point as $\gamma_0''$.  $r(\gamma)$ is convex in $[0,\gamma_0'')$ and concave in $[\gamma_0'',+\infty)$. Therefore, whether $r(\gamma)$ is convex in $[\gamma_0,+\infty)$ is determined by the relationship between $\gamma_0$ and $\gamma_0''$.
Since $\gamma_0$ is the zero point of $r(\gamma)$, we have
\begin{equation}
\begin{aligned}
\log_2(\gamma_0+1)&-\log_2(e)\eta\sqrt{1-\frac{1}{(\gamma_0+1)^2}}=0\\
&\Rightarrow   \eta=\frac{(\gamma_0+1)\ln(\gamma_0+1)}{\sqrt{\gamma_0^2+2\gamma_0}}.
\end{aligned}
\end{equation}
On this basis, an auxiliary function is introduced as follows
\begin{equation}
\label{f20}
f_1(\gamma)=\frac{(\gamma+1)\ln(\gamma+1)}{\sqrt{\gamma^2+2\gamma}}.
\end{equation}
Similarly, since $\gamma_0''$ is the zero point of $\bigtriangledown^2 r(\gamma)$, we have that
\begin{equation}
\begin{aligned}
\frac{2\eta}{(\gamma_0''+1)([\gamma_0'']^2+2\gamma_0'')^{\frac{1}{2}}}+ \frac{\eta(\gamma_0''+1)}{([\gamma_0'']^2+2\gamma_0'')^\frac{3}{2}}-1=0 &\\
\Rightarrow \quad  \eta=\frac{(\gamma_0''+1)(\gamma_0''^2+2\gamma''_0)^{\frac{3}{2}}}{3\gamma_0''^2+6\gamma''_0+1}&.
\end{aligned}
\end{equation}
Another auxiliary function is introduced as follows
\begin{equation}
f_2(\gamma)=\frac{(\gamma+1)(\gamma^2+2\gamma)^{\frac{3}{2}}}{3\gamma^2+6\gamma+1}.
\end{equation}
The fact that there is only one zero point of $r(\gamma)$ (except 0) and $\triangledown^2 r(\gamma)$ means that, for any $\eta>0$, $\gamma_0$ and $\gamma_0''$ are the only solutions to $f_1(\gamma)=\eta$ and $f_2(\gamma)=\eta$. In other words, $f_1(\gamma)$ and $f_2(\gamma)$ are injective functions when $\gamma>0$. Therefore, we can turn the comparison of $\gamma_0$ and $\gamma_0''$ into the comparison of $f_1(\gamma)$ and $f_2(\gamma)$.
If $\eta$ is in the region of $f_1(\gamma)>f_2(\gamma)$, $\gamma_0<\gamma_0''$ holds, otherwise, $\gamma_0\geqslant\gamma_0''$ holds.
Then,  we denote $f(\gamma)=f_1(\gamma)-f_2(\gamma)$
\begin{equation}
f(\gamma)=\frac{(\gamma+1)}{\sqrt{\gamma^2+2\gamma}}\bigg[\mbox{In}(\gamma+1)-\frac{(\gamma+1)(\gamma^2+2\gamma)^2}{3\gamma^2+6\gamma+1}\bigg]. \nonumber
\end{equation}
It can be further proven that $f(\gamma)$ is first positive and then negative in $[0,+\infty)$.
Denoting its zero-crossing point as $\hat{\gamma}$, we could draw the conclusion that
\begin{equation}
\label{f19}
\left\{
\begin{aligned}
f_1(\gamma)\geqslant f_2(\gamma), \ \ &\gamma \leqslant\hat{\gamma}, \\
%	f_1(\gamma)=f_2(\gamma), \ \ &\gamma=\hat{\gamma}, \\
f_1(\gamma)<f_2(\gamma), \ \ &\gamma>\hat{\gamma}.
\end{aligned}
\right.
\end{equation}
Then, given the value of $\eta$, the relationship between $\gamma_0$ and $\gamma_0''$ can be judged by determining whether $\eta$ is greater than $f_1(\hat{\gamma})$ (or $f_2(\hat{\gamma})$)
\begin{equation}
\left\{
\begin{aligned}
\gamma_0\leqslant\gamma_0'', \ \ &\eta\leqslant f_1(\hat{\gamma}), \\
\gamma_0>\gamma_0'', \ \ &\eta> f_1(\hat{\gamma}).
\end{aligned}
\right.
\end{equation}
Thus, we can further draw the conclusion that
\begin{equation}
\left\{
\begin{aligned}
&r(\gamma) \  \text{is convex-concave,}  \quad \eta\leqslant\frac{(\hat{\gamma}+1)\ln(\hat{\gamma}+1)}{\sqrt{\hat{\gamma}^2+2\hat{\gamma}}}, \\
&r(\gamma) \ \text{is concave,}  \qquad\qquad\quad \eta> \frac{(\hat{\gamma}+1)\ln(\hat{\gamma}+1)}{\sqrt{\hat{\gamma}^2+2\hat{\gamma}}}.
\end{aligned} \blacksquare
\nonumber
\right.
\end{equation}

\section{Proof of \bf{Theorem 2}}
For simplicity, we rewrite the cycle rate and the rate-cost function as the functions of the SNR, i.e., $R(\gamma)$ and $L(\gamma)$. %Other parameters are assumed to be given in advance.
In addition, a new function is introduced as follows
\begin{equation}
\label{o}
\Omega(\gamma)=R(\gamma)-\log_2|\det\mathbf{A}|.
\end{equation}
To figure out the convexity of $L(\gamma)$, we calculate its second derivative
\begin{small}
\begin{equation}
\label{diff3}
\begin{aligned}
\triangledown^2 L(\gamma)=
&\frac{2^{\frac{2}{m}\Omega(\gamma)+1}|\det \mathbf{N}\mathbf{M}|^\frac{1}{m}}{\log_2e(2^{\frac{2}{m}\Omega(\gamma)}-1)^2}\\
&\times\bigg[\frac{2}{m\log_2e}\big(\frac{2^{\frac{2}{m}\Omega(\gamma)}+1}{2^{\frac{2}{m}\Omega(\gamma)}-1}\big)[\triangledown \Omega(\gamma)]^2-\triangledown^2 \Omega(\gamma)\bigg],
\end{aligned}
\end{equation}
\end{small}
where $\triangledown \Omega(\gamma)$ and $\triangledown^2 \Omega(\gamma)$ are given by
\begin{equation}
\label{do}
\triangledown \Omega(\gamma)=\log_2e\big(\frac{n}{\gamma+1}-\frac{\sqrt{n}Q^{-1}(\epsilon)}{(\gamma+1)^2\sqrt{\gamma^2+2\gamma}}\big),
\end{equation}
\begin{equation}
\label{ddo}
\begin{aligned}
\triangledown^2 \Omega(\gamma)=\log_2e\bigg(&-\frac{n}{(\gamma+1)^2}\\
&+\sqrt{n}Q^{-1}(\epsilon)\frac{(\gamma+1)^2+2(\gamma^2+2\gamma)}{(\gamma+1)^3(\gamma^2+2\gamma)^{\frac{3}{2}}}\bigg). %\label{diff_o}
\end{aligned}
\end{equation}
%To prove the convexity of $L(\gamma)$, we shall show that $\triangledown^2 L(\gamma)\geqslant0$ holds for any $m$.
Accordingly, we define two functions as follows:
\begin{equation}
\label{f}
f(m,\gamma)\triangleq\frac{2}{m\log_2e}\big(\frac{2^{\frac{2}{m}\Omega(\gamma)}+1}{2^{\frac{2}{m}\Omega(\gamma)}-1}\big)[\triangledown \Omega(\gamma)]^2-\triangledown^2 \Omega(\gamma),
\end{equation}
\begin{equation}
g(m,\Omega)\triangleq\frac{2}{m\log_2e}\big(\frac{2^{\frac{2}{m}\Omega}+1}{2^{\frac{2}{m}\Omega}-1}\big).
\end{equation}
We first prove that $g(m,\Omega)$ is monotonically decreasing with $m$. Its partial derivative of $m$ is given by
\begin{equation}
\frac{\partial g(m,\Omega)}{\partial m}=\frac{2}{\log_2(e)m^2(2^{\frac{2\Omega}{m}}-1)^2}\big[\frac{\Omega2^{\frac{2\Omega}{m}+2}}{m\log_2e}-2^{\frac{4\Omega}{m}}+1\big].
\end{equation}
The sign of $\frac{\partial g(m,\Omega)}{\partial m}$ is determined by the term  $\big[\frac{\Omega2^{\frac{2\Omega}{m}+2}}{m\log_2e}-2^{\frac{4\Omega}{m}}+1\big]$. It is obvious that
\begin{equation}
\label{1}
\frac{\Omega2^{\frac{2\Omega}{m}+2}}{m\log_2e}-2^{\frac{4\Omega}{m}}+1|_{\Omega=0}=0.
\end{equation}
In addition, we calculate the partial derivative of the above expression of  $\Omega$. The expression is given by
\begin{equation}
\frac{2^{\frac{2\Omega}{m}+2}}{m\log_2e}\bigg(\frac{2\Omega}{m\log_2e}+1-2^{\frac{2\Omega}{m}}\bigg).
\end{equation}
Since $e^x\geqslant x+1$, we could find that $\frac{2\Omega}{m\log_2e}+1-2^{\frac{2\Omega}{m}}\leqslant 0$. Based on \eqref{1}, it is easy to  draw the conclusion that
\begin{equation}
\frac{\partial g(m,\Omega)}{\partial m}|_{\Omega \geqslant 0} \leqslant 0.
\end{equation}
Therefore, $g(m,\Omega)$ is monotonically decreasing with $m$. The minimal value of $f(m,\gamma)$ is obtained when $m$  goes to infinity. If we could prove $\lim\limits_{m\rightarrow \infty}f(m,\gamma)\geqslant 0$, the non-negativity of $\triangledown^2 L(\gamma)$ can be proven.

In the following,  we consider the limit case of  $m\rightarrow +\infty$. In this case, the index number $\frac{2\Omega}{m}$ in $g(m,\Omega)$ can be arbitrarily small. To facilitate the simplification, we assume  $\frac{2\Omega}{m}  \leqslant 1$. Then, it is easy to have the following scaling
\begin{equation}
\frac{2\Omega}{m}  \leqslant 1 \Rightarrow 2^{\frac{2}{m}\Omega}-1\leqslant \frac{4\Omega}{m\log_2e}.
\end{equation}
On this basis, we can further scale $g(m,\Omega)$ as
\begin{small}
\begin{equation}
\label{2}
\begin{aligned}
&\lim\limits_{m \rightarrow +\infty}g(m,\Omega)=\lim\limits_{m \rightarrow +\infty}\frac{2}{m\log_2e}\big(1+\frac{2}{2^{\frac{2}{m}\Omega}-1}\big)\\
&\geqslant \frac{4}{m\log_2e(2^{\frac{2}{m}\Omega}-1)}
\geqslant\frac{4}{m\log_2e(\frac{4\Omega}{m\log_2e})}=\frac{1}{\Omega}
\end{aligned}
\end{equation}
\end{small}
Moving forward, by substituting \eqref{2} into \eqref{f}, we have
\begin{equation}
\label{o4}
\lim\limits_{m \rightarrow +\infty} f(m,\gamma)\geqslant \frac{1}{\Omega(\gamma)}[\triangledown \Omega(\gamma)]^2-\triangledown^2 \Omega(\gamma).
\end{equation}
%Accordingly, we define $\mathcal{F}(\gamma)\triangleq\frac{1}{\Omega(\gamma)}[\triangledown \Omega(\gamma)]^2-\triangledown^2 \Omega(\gamma)$.
Furthermore, $\Omega(\gamma)$ can be scaled into
\begin{equation}
\label{o1}
\Omega(\gamma) \leqslant n\log_2(\gamma+1) \leqslant  n\gamma \log_2e.
\end{equation}
In addition, as we mentioned after \eqref{e1}, $L(\gamma)$ makes sense only when the stable condition is ensured,
\begin{equation}
\begin{aligned}
R(\gamma)-\log_2|\det\mathbf{A}|  &\geqslant 0\\
\Rightarrow R(\gamma)	&\geqslant0\\
\Rightarrow Q^{-1}(\epsilon)&< \frac{\sqrt{n}\log_2(\gamma+1)}{\log_2e\sqrt{1-\frac{1}{(\gamma+1)^2}}}. \label{Q}
\end{aligned}
\end{equation}
By substituting $Q^{-1}(\epsilon)$ with the right side of  \eqref{Q},  we could derive the following inequality of $\triangledown \Omega(\gamma)$ \eqref{do} and $\triangledown^2 \Omega(\gamma)$ \eqref{ddo}
%based on the inequality \eqref{Q}, $\triangledown \Omega(\gamma)$ and $\triangledown^2 \Omega(\gamma)$ can be relaxed as
\begin{equation}
\label{o2}
\begin{aligned}
\triangledown \Omega(\gamma) &\geqslant n\log_2e\big(\frac{1}{\gamma+1}-\frac{\log_2(\gamma+1)}{\log_2e(\gamma+1)(\gamma^2+2\gamma)}\big)\\
&\geqslant n\log_2e\big(\frac{1}{\gamma+1}-\frac{1}{(\gamma+1)(\gamma+2)}\big),
\end{aligned}
\end{equation}
\begin{equation}
\label{o3}
\begin{aligned}
\triangledown^2 \Omega(\gamma) &\leqslant n\log_2e\bigg(-\frac{1}{(\gamma+1)^2}+ \frac{\log_2(\gamma+1)}{\log_2e}*\\
&\quad\quad\quad\quad\quad\quad\frac{(\gamma+1)^2+2(\gamma^2+2\gamma)}{(\gamma+1)^2(\gamma^2+2\gamma)^{2}}\bigg)\\
&\leqslant n\log_2e\bigg(-\frac{1}{(\gamma+1)^2}+\frac{(\gamma+1)^2+2(\gamma^2+2\gamma)}{\gamma(\gamma+1)^2(\gamma+2)^2}\bigg).
\end{aligned}
\end{equation}
By substituting \eqref{o1}, \eqref{o2} and \eqref{o3} into \eqref{o4}, we have that
\begin{equation}
\begin{aligned}
&\lim\limits_{m \rightarrow +\infty} f(m,\gamma)\geqslant \\ &\frac{1}{n\gamma\log_2e}\bigg(n\log_2e\big(\frac{1}{\gamma+1}-\frac{1}{(\gamma+1)(\gamma+2)}\big)\bigg)^2-\\
&n\log_2e\bigg(-\frac{1}{(\gamma+1)^2}+\frac{(\gamma+1)^2+2(\gamma^2+2\gamma)}{\gamma(\gamma+1)^2(\gamma+2)^2}\bigg)\\
&\quad \quad \quad  \qquad \quad=\frac{n\gamma\log_2e}{(\gamma+1)^2(\gamma+2)} \geqslant 0.
\end{aligned}
\end{equation}
Therefore, for any $m$, we have that
\begin{equation}
\triangledown^2 L(\gamma) \geqslant\lim\limits_{m \rightarrow +\infty} f(m,\gamma)\geqslant 0.
\end{equation}
The convexity of $L(\gamma)$ is thus proven. $\blacksquare$

\end{document}